\documentclass[reprint,superscriptaddress,aps,pre,floatfix]{revtex4-2}
\usepackage{microtype}
\usepackage[utf8]{inputenc}
\usepackage[tbtags]{amsmath}
\allowdisplaybreaks
\usepackage{amsthm,amssymb,dsfont,mathrsfs,bbm,xfrac,tabularx,booktabs}
\usepackage[linesnumbered,ruled,vlined]{algorithm2e}
\usepackage{upgreek}
\usepackage{mathtools}

\newtheorem*{prop*}{Proposition}

\usepackage{comment}
\usepackage{tikz}
\usetikzlibrary{patterns,decorations,arrows,shapes.geometric,arrows,mindmap,backgrounds,positioning,positioning,fit,backgrounds,positioning,arrows,matrix,calc}
\usetikzlibrary{shapes,backgrounds,mindmap,shadows,shapes.geometric,topaths,arrows,pgfplots.groupplots,fadings,calc,decorations.markings,positioning,chains,fit}
\tikzfading[name=fade l,left color=transparent!100,right color=transparent!0]
\tikzfading[name=fade r,right color=transparent!100,left color=transparent!0]
\tikzfading[name=fade d,bottom color=transparent!100,top color=transparent!0]
\tikzfading[name=fade u,top color=transparent!100,bottom color=transparent!0]
\usepackage{braket}
\usepackage{pgfplotstable}
\usepackage{enumitem}

\usepackage{ifthen}

\pgfplotsset{compat=1.16}

\newdimen\nodeSize
\newdimen\nodeDist
\tikzset{position/.style args={#1:#2 from #3}{at=(#3.#1), anchor=#1+180, shift=(#1:#2)}}

\newcommand{\beq}{\begin{equation}}
\newcommand{\eeq}{\end{equation}}
\newcommand{\beqa}{\begin{eqnarray}}
\newcommand{\eeqa}{\end{eqnarray}}

\newcommand{\<}{\langle}
\renewcommand{\>}{\rangle}

\def\<{\langle}
\def\>{\rangle}

\begin{document}

\title{Solution of the Critical Dynamics of the Mean-Field Kob-Andersen Model}

\author{Gianmarco Perrupato}
\affiliation{Department of Computing Sciences, Bocconi University, 20136 Milano, Italy}
\author{Tommaso Rizzo}
\affiliation{Institute of Complex Systems (ISC) - CNR, Rome unit, P.le A. Moro 5, 00185 Rome, Italy}
\affiliation{Dipartimento di Fisica, Sapienza Universit\`a di Roma, P.le A. Moro 5, 00185 Rome, Italy}
\begin{abstract}

We analytically solve the critical dynamics of the Kob-Andersen kinetically constrained model of supercooled liquids on the Bethe lattice, employing a combinatorial argument based on the cavity method. For arbitrary values of graph connectivity $z$ and facilitation parameter $m$, we demonstrate that the critical behavior of the order parameter is governed by equations of motion equivalent to those found in Mode-Coupling Theory. The resulting predictions for the dynamical exponents are validated through direct comparisons with numerical simulations that include both continuous and discontinuous transition scenarios.
\end{abstract}

\maketitle
\section{introduction}
Understanding the dynamics of supercooled liquids remains a fundamental open challenge in statistical physics. A cornerstone of this quest is represented by the analytical study of certain mean-field models of spin-glasses \cite{kirkpatrick1989scaling,thirumalai1988mean,kirkpatrick1989random} and simple liquid models in the limit of infinite dimensions \cite{charbonneau2017glass}, both described by Parisi's replica-symmetry-breaking (RSB) theory. These RSB models of glasses (RSBGs) share complex thermodynamics properties, and exhibit many non-trivial dynamical features, e.g.\ the existence of an ergodicity-breaking transition characterized by a two timescales relaxation of the order parameter close to the critical point. Moreover, their dynamics is described by a set of equations
that are formally the same as the approximated equations formulated by Mode Coupling Theory (MCT) for finite-dimensional structural glasses \cite{gotze2008complex}. 

Remarkably, this phenomenology appears to extend beyond RSBGs. Indeed the dynamics of certain kinetically constrained models (KCMs)  exhibit qualitatively similar behavior \cite{kurchan1997aging}, 
in particular on mean-field geometries called Bethe lattices (BLs) \cite{sellitto2005facilitated,sellitto2015,de2016scaling,sellitto2010dynamic,franz2013finite,ikeda2015fredrickson,ikeda2017fredrickson}. See \cite{ritort2003glassy,garrahan2011kinetically,HartarskyToninelli2025}  for broad reviews on KCMs. BLs are finite-connectivity random graphs enjoying the so-called locally-tree-likeness, i.e.\ the neighborhood of a node taken at random is typically a tree (namely there are no loops) up to a distance that scales like $\log{N}$, where $N$ is the size of the system. This topological property, combined with a probabilistic argument, has been demonstrated to suffice for analytically solving the dynamics of a family of KCMs on BLs, known as Fredrickson-Andersen models (FA) \cite{perrupato2025theory,perrupato2024thermodynamics}. The solution of the dynamics of FA reveals that their critical behavior is governed by the same MCT equations as those found in RSBGs, as previously observed also in numerical simulations \cite{toninelli2005cooperative,sellitto2015}. This observation is particularly striking, given the fundamental differences between these two classes of models. Unlike RSBGs, where complex behavior arises from thermodynamic singularities, in KCMs it emerges from a purely dynamical mechanism known as \emph{facilitation} (which we will discuss later). As a result, KCMs are often presented as an extreme case in the debate over the origins of glassy behavior—specifically, whether glassiness stems from thermodynamic causes or is purely a dynamical phenomenon, a question central to the so-called dynamics versus structure dilemma.

Another central problem is the comparison between the predictions obtained by the analytical study of the aforementioned mean-field models, and numerical simulations of supercooled liquids in finite dimensions. Indeed, the dynamical arrest transition characterizing both RSBGs and KCMs on BLs is not found in numerical experiments of finite-dimensional models of glasses, but rather a crossover from power-law to exponential increase of the relaxation time. The possibility is that such ``spurious'' transition is a consequence of the mean-field nature of the models. Perturbations around mean-field theory can be taken into account by a renormalization-group approach, which in spin-glasses leads to a set of stochastic dynamical equations for the order parameter called Stochastic-Beta-Relaxation (SBR) equations \cite{rizzo2016dynamical,rizzo2022stochastic}. The same program can be carried out also for FA \cite{rizzo2020solvable}. In both cases it is found that within SBR the arrest transition is turned into a crossover, as observed in realistic systems. An important observation is that given the analytical solution of the dynamics at the mean-field level, the SBR equations lead to parameter-free predictions beyond mean-field. This is one of the key reasons motivating the study of the dynamics of MF models, that in the case of KCMs has been solved only for FA. 

In this paper we analytically solve the critical dynamics of another prototypical family of KCMs on the BL called Kob-Andersen models (KA) \cite{kob1993kinetic}. KA is a lattice gas model that, at variance with FA, has conserved dynamics, namely the number of particles is a constant of motion. We show that the argument used in \cite{perrupato2025theory} to solve FA can be extended also to the KA case, leading to an equation of motion for the order parameter close to the critical point equivalent to those found in Mode-Coupling Theory. The paper is organized as follows. In Sec.~\ref{sec:defKA}, we define KA on the Bethe lattice. In particular, in Subsec.~\ref{sec:defKAglassPhen} we introduce the glassy phenomenology of the model. In Subsec.~\ref{sec:CavityKA} we discuss the cavity method for the computation of the plateau value of the persistence, and the critical point. In Sec.~\ref{sec:dynKAcav}, we derive an exact closed equation of motion for the order parameter of the problem, the persistence function, in the $\beta$-regime. Specifically, in Subsec.~\ref{ssec:theblockedpers} we define the fundamental objects required for the study of the critical behaviour of the persistence. In Subsec.~\ref{ssec:critHier} we present the argument for the derivation of the equation of motion, and discuss the case of KA with discontinuous transitions. In Subsec.~\ref{ssec:ContinuousModelsKA} we study the case of KA with continuous transition. Finally, in Sec.~\ref{sec:concl} we present the conclusions.

\section{Kob-Andersen Model}
\label{sec:defKA}

\subsection{Definitions and glassy phenomenology}
\label{sec:defKAglassPhen}

In KA each site of the graph is allowed to be occupied by zero or one particle. We call $N$ the number of sites. The particles move according to a kinetically constrained dynamics that conserves their number and is defined as follows. An initial configuration is generated by independently populating each site with probability $\rho$, where $\rho$ is the density of the particle number. At each time-step, a randomly chosen particle attempts to move to one of its neighbors, chosen at random. However, a move from site $i$ to site $j$ is allowed if and only if: i) site $j$ is empty, ii) the particle has no more than $m$ occupied neighbors before \emph{and} after the move, where $m$ is a \emph{facilitation} parameter. Within a given configuration, we define facilitated sites as either occupied sites where the particle is allowed to move, or empty sites that can be occupied. Conversely, blocked sites are those that are not facilitated. Note that the above dynamics satisfies the detail balance condition with respect to the probability distribution of the initial configuration, which is factorized over the sites. However, the facilitation constraints introduce non-trivial properties into the model, which depend on the underlying graph topology. An example is provided by BLs with fixed coordination $z$ (i.e.\ each site has $z$ neighbors), which are equivalent in the large-$N$ limit to random regular graphs (RRGs). In this case, for $0<m<z-1$, there exists a critical density $0<\rho_c(z,m)<1$ such that the KA dynamics is ergodic for $\rho<\rho_c(z,m)$, while ergodicity breaking occurs for $\rho\geq \rho_c(z,m)$ \cite{toninelli2005cooperative}. In \cite{toninelli2005cooperative,cancrini2010kinetically} it is proven that on hyper-cubic lattices of dimension $d$, there is no ergodicity breaking for $d\leq m\leq 2d-2$, and for $m<d$ there is a finite fraction of frozen particles at any density. Note that the case $m=2d-1$, corresponding to an unconstrained lattice gas, is trivial. 

\begin{figure}
\centering
\includegraphics[width=0.98\columnwidth]{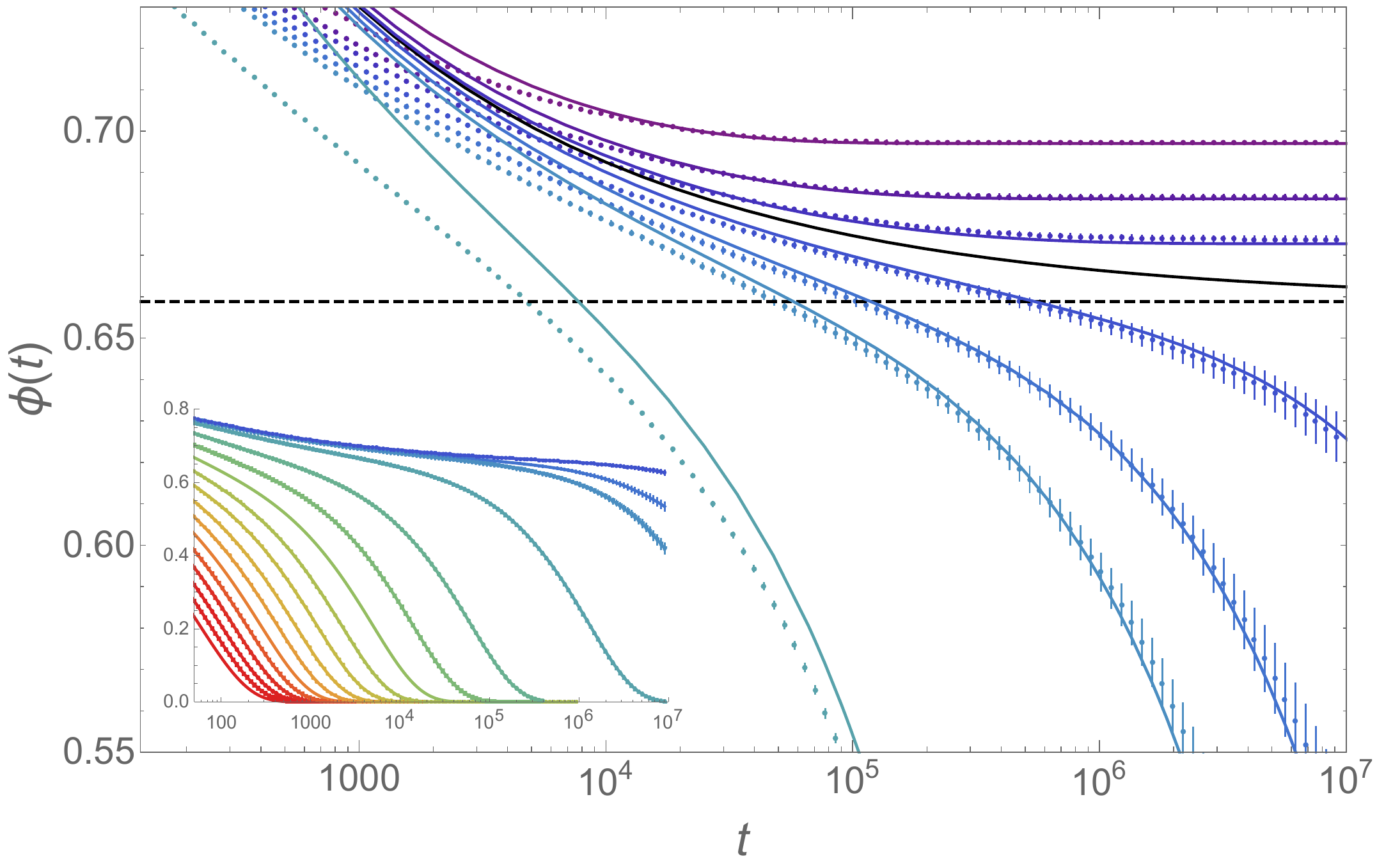}
\caption{Double-step relaxation of the persistence of KA on RRGs with $z=6$, $m=3$. Points correspond to numerical simulations of systems with size $N=64\times 10^5$. From bottom left to top right the densities are $\rho= 0.83, 0.8335, 0.834, 0.8345, 0.835, 0.8355, 0.8365$. Continuous lines represent the numerical solution of the MCT integral equation \eqref{MCTcrit}, with $\lambda$ and $\sigma$ computed analytically. In particular, $\sigma=c_{z,m}\,(\rho-\rho_c)$, where the constant $c_{z,m}$ is predicted by our theory (see App.~\ref{app:tdyneq}). For $z=6$, $m=3$, $c_{6,3}\approx 0.251$. Note that the microscopic timescale $t_0$ appearing in the initial condition of Eq.~\eqref{MCTcrit} is not fixed by $\lambda$ and $c_{z,m}$. We fit $t_0$ from the data at the critical point, where $\phi(t)\approx\phi_{plat}+(t/t_0)^{-a}$, obtaining $t_0\approx 0.3$. \emph{Inset}: persistence versus time for various densities ($0.7\leq\rho\leq 0.8345$).\label{fig:KApersistence}}
\end{figure}

On the BL the critical dynamics of KA is qualitatively the same as that of other KCMs, like FA. An order parameter of the problem is provided by the persistence function $\phi(t)$, which counts the fraction of sites occupied by a particle that never moved up to time $t$. The ergodicity breaking transition $\rho=\rho_c(z,m)$, separates a ``low-density'' (liquid) phase $\rho<\rho_c(z,m)$ where the persistence $\phi(t)$ relaxes to zero in the long-time limit, from a ``high-density'' (glassy) phase $\rho>\rho_c(z,m)$ where $\phi(t)$ reaches a non-zero plateau (see Fig.~\ref{fig:KApersistence}). This implies that, in the glassy phase, a finite fraction of particles are \emph{frozen}, i.e.\ no sequence of dynamical updates can ever change their position. Depending on the facilitation value $m$, the transition at $\rho_c(z,m)$ is discontinuous for $1<m<z-1$ or continuous for $m=1$ \cite{toninelli2005cooperative}. We call $\phi_{plat}(z,m)$ the plateau value of the persistence function at the critical point; in the following, the dependence of $\phi_{plat}$ and $\rho_c$ on $z$ and $m$ will be omitted when not explicitly needed. 

As we show in the next sections, the relaxation of the persistence function close to the critical point in the liquid phase is characterized by two timescales described by MCT dynamical equations. This means that for densities close to the critical one, $\rho\lesssim \rho_c$, there is a so-called $\beta$-regime corresponding to a timescale $\tau_\beta$ on which the persistence is almost equal to $\phi_{plat}$ followed by a so-called $\alpha$-regime, during which the persistence decays from $\phi_{plat}$ to zero. 

Within MCT the deviation $g(t)$ of the dynamical correlator from its plateau value in the $\beta$ regime is controlled by the following equation \cite{gotze2008complex}:
\beq
\label{MCTcrit}
\sigma = - \lambda \, g^2(t) + \frac{d}{dt} \, \int_0^t  \, g(t')\, g(t-t')\, dt' \ ,
\eeq
where $\sigma$ is a linear function of the distance of the control parameter (e.g.\ temperature or particle density) from its critical value, and $\lambda$ is a model-dependent constant known as the parameter exponent. Equation \eqref{MCTcrit} has to be solved with the initial condition $\lim_{t\to 0} g(t)\,(t/t_0)^a = 1$, where the critical exponent $a$ is uniquely determined by $\lambda$ (see below), and $t_0$ sets the microscopic timescale of the dynamics. At the critical point ($\sigma=0$), $g(t)$ goes to zero with the power-law behavior $g(t)=(t/t_0)^{-a}$, and close to criticality it obeys the following scaling laws. To describe them, we introduce the master functions $f_{\pm}$, which are the solutions of Eq.~\eqref{MCTcrit} with $\sigma=\pm 1$, $t_0=1$. In the $\beta$-regime, $g(t)$ can be expressed as:
\begin{equation}
\label{eq:scalinglaw}
g(t) = |\sigma|^{1/2} \, f_{\pm}(t/\tau_\beta),
\end{equation}
where $\tau_\beta=t_0\, |\sigma|^{-1/(2\,a)}$ defines the $\beta$ timescale, and $f_+$ or $f_-$ are chosen depending on whether $\sigma$ is positive or negative, respectively. The function $f_+(t)$ describes the glassy regime ($\sigma>0$) close to the critical point, where the dynamical correlator remains blocked for $t\to\infty$. In particular $\lim_{t\rightarrow\infty}f_{+}(t)=(1-\lambda)^{-1/2}$. The function $f_{-}(t)$ describes the critical behavior in the liquid phase ($\sigma<0$). In this regime, the correlator departs from the plateau according to the von Schweidler law $f_{-}(t)\propto -(t/\tau_{\beta})^{b}$. The critical exponents $a$ and $b$ are determined by the parameter exponent $\lambda$ as follows: 
\beq
\label{lambdaMCT}
\lambda=\frac{\Gamma^2(1-a)}{\Gamma(1-2a)}=\frac{\Gamma^2(1+b)}{\Gamma(1+2b)}.
\eeq
Equation \eqref{MCTcrit} is valid provided that $g(t)$ is small, and this condition defines the $\beta$-regime. However, from the equation itself one can also deduce the timescale of the $\alpha$ regime, $\tau_{\alpha}$, defined as the characteristic time at which the deviation from the plateau decays to a constant value independent of $\sigma$. According to the von Schweidler law, this condition corresponds to $(\tau_{\alpha}/\tau_{\beta})^{b}\,|\sigma|^{1/2}=O(1)$, which implies the divergence $\tau_\alpha \propto |\sigma|^{-\gamma}$ as $\sigma \to 0$, with the critical exponent $\gamma=1/(2a)+1/(2b)$.  

All of the above scaling laws have been shown to hold in FA with $g(t)\equiv \phi(t)-\phi_{plat}$ \cite{perrupato2025theory}, where $\phi(t)$ is a persistence function analogous to the one we defined for KA, and $\phi_{plat}$ is its plateau value at the critical point. In the following, we show that the same analysis can be extended to KA, deriving an equation for the deviation of the persistence from its plateau value equivalent to \eqref{MCTcrit} together with analytical expressions for $\lambda$ and $\sigma$.

\subsection{Cavity equations for the plateau and the critical density}
\label{sec:CavityKA}
The plateau value of the persistence and the critical density can be computed on the BL with the cavity method, analogously to the FA case. In the following, we focus on BLs with fixed coordination $z$. Based on \cite{toninelli2005cooperative}, we define the following ``cavity'' probabilities (or cavity persistences) $Y,B$ and $F$: i) $Y$ is the probability that a site is occupied by a frozen particle, conditioned to one of its neighbors to be occupied; ii) $B$ is the probability that a site can never be occupied by a particle, conditioned to one of its neighbors to be occupied; iii) $F$ is the probability that a site is occupied by a frozen particle, conditioned to one of its neighbors to be empty. 

The plateau value $\phi_{plat}$ of the persistence $\phi(t)$, which corresponds to the probability that a site on the graph is occupied by a frozen particle, can be easily expressed in terms of $Y$ and $B$:
\begin{multline}
\label{eq:defphiKA}
\phi_{plat} = \rho \sum_{j=m+1}^z \binom{z}{j} Y^j (1-Y)^{z-j} +\\+\rho \sum_{j=0}^m \binom{z}{j} Y^j B^{z-j}.
\end{multline}
The cavity probabilities $Y$, $B$, and $F$ are obtained as solutions of the self-consistent equations
\begin{equation}
\label{eq:KAcavY}
Y = \rho\, G^{(Y)}(Y,B),
\end{equation}
\begin{equation}
\label{eq:KAcavB}
B = (1-\rho)\, G^{(B)}(F),
\end{equation}
\begin{equation}
\label{eq:KAcavF}
F = \rho\, G^{(F)}(Y,B),
\end{equation}
with the explicit expressions provided in Appendix~\ref{app:cavityplat}. We now summarize the main properties of this system (see \cite{toninelli2005cooperative}). For all densities $\rho$, the liquid state, characterized by $Y=B=F=0$ (equivalently $\phi_{\text{plat}}=0$), is a fixed point. At small $\rho$ this is the only solution, while for sufficiently large $\rho$ the system also admits a ``glassy'' solution with non-zero cavity persistences. The nature of the transition point, where this new solution appears, depends on the value of $m$. As anticipated in Sec.~\ref{sec:defKAglassPhen}, for $m=1$, the transition is continuous, namely, by increasing $\rho$ the glassy fixed point detaches continuously from the liquid fixed point, which becomes unstable above $\rho_c$. In this case $\rho_c=1/(z-1)$. For $m>1$, the liquid fixed point remains always stable and the transition is discontinuous, that is, the glassy fixed point is created away from the liquid one: upon increasing $\rho$ the cavity persistences jump discontinuously at $\rho_c$, where $\phi_{plat}(\rho_c)>0$. The critical values of $Y,B,F$ and $\rho$ are given by the solution of Eqs.~\eqref{eq:KAcavY}, \eqref{eq:KAcavB} and \eqref{eq:KAcavF}, with the additional request that the linearized system admits an eigenvalue equal to one (see Appendix~\ref{app:cavityplat}):
\begin{equation}
\label{eq:CritConddet}
\det{\big(\mathds{1}-\boldsymbol{J}(\boldsymbol{P},\rho)\big)}=0,
\end{equation}
where $\mathbf{P}=(Y,B,F)$ is the cavity persistence vector, $\mathds{1}$ is the $3\times 3$ identity matrix, and $\boldsymbol{J}$ is the Jacobian matrix obtained differentiating $\rho\,G^{(Y)}$, $(1-\rho)\,G^{(B)}$, and $\rho\,G^{(F)}$ with respect to the cavity persistences $Y$, $B$, and $F$. 

We note that the frozen cluster depends on the initial configuration of the dynamics. A natural question is whether it is uniquely fixed by this configuration, or if it might also depend on the specific dynamical history, namely, whether different sequences of moves could lead to different frozen clusters. One can show that any such dependence on the dynamics leads to a contradiction. Suppose that there is a sequence of moves, call it $A$, such that a given particle will be frozen for all times after sequence $A$ is completed, while it can be moved through another sequence, say $B$. Now after sequence $A$ is completed, one can invoke the reversibility of the dynamics to go back to the initial condition and then perform sequence $B$, in contradiction with the assumption that the particle could not be moved after sequence $A$. Hence, the frozen cluster is uniquely determined by the initial configuration. 

An explicit procedure involving a bootstrap percolation (BP) process to identify particles belonging to the frozen cluster was given in \cite{toninelli2005cooperative}. In this process, starting from the initial condition, particles that can move according to the KA rules are culled. This violates the conservation of the particle number of the model and, as a consequence, the frozen cluster of the KA model in general is larger than the one of the BP process. However, it is argued that particle conservation is only essential in the presence of finite loops in the graph, and therefore it is irrelevant on the BL. Indeed, on the BL the BP process is described by the very same equations \eqref{eq:KAcavY}, \eqref{eq:KAcavB} and \eqref{eq:KAcavF} of the KA model.

\section{Dynamical equations in the $\beta$ regime}
\label{sec:dynKAcav}

\subsection{Blocked persistences}
\label{ssec:theblockedpers}

In order to derive the equations for the dynamics of the persistence, we start with a number of definitions. 

We define the time-dependent counterparts $Y(t), B(t)$, and $F(t)$ of the cavity persistences introduced in Sec.~\ref{sec:CavityKA} as follows: i) $Y(t)$ is the probability that a site is occupied by a particle that did not move up to time $t$, conditioned to one of its neighbors to be occupied; ii) $B(t)$ is the probability that a site remains empty up to time $t$, conditioned to one of its neighbors to be occupied; iii) $F(t)$ is the probability that a site is occupied by a particle that did not move up to time $t$, conditioned to one of its neighbors to be empty. For $t\to\infty$, $Y(t),B(t)$ and $F(t)$ tend respectively to the cavity persistences $Y,B$ and $F$. In what follows, we adopt the vector notation
\begin{equation}
\boldsymbol{\delta \hat{\phi}}(t)=\big(\delta Y(t),\,\delta B(t),\,\delta F(t)\big),
\end{equation}
and we use the symbol ``$\delta$'' in front of a persistence to denote the persistence minus its plateau value. 

Following \cite{perrupato2025theory}, we also introduce the {\it blocked persistence} $\phi_b(t)$ as the fraction of particles that have been {\it blocked} at all times less than $t$. We define analogously the blocked cavity persistences $Y_b(t), B_b(t), F_b(t)$, and we use the notation:
\begin{equation}
\boldsymbol{\delta\hat{\phi}}_b(t)=\big(\delta Y_b(t),\,\delta B_b(t),\,\delta F_b(t)\big).
\end{equation}
By definition, $\phi(t) \geq \phi_b(t)$, since a facilitated particle (i.e.\ not blocked) does not necessarily move. In the infinite-time limit both quantities converge to the same plateau, $\phi_{plat}$. Crucially, at large times, the deviations $\delta \phi(t)$ and $\delta \phi_b(t)$ vanish with the same leading behavior at the critical point, and the same holds for the blocked cavity persistences, $\boldsymbol{\delta\hat{\phi}}_b(t)\approx \boldsymbol{\delta\hat{\phi}}(t)$ (see App.~\ref{app:EqPersBlockPers} and \cite{perrupato2025theory}). 

Keeping in mind the definitions given in Sec.~\ref{sec:CavityKA}, we note that for a particle $i$ to be blocked at a certain time, it must have a \emph{blocking set}, i.e.\ a set composed by either: a) at least $m+1$ neighboring occupied sites, or b) at least $z-m$ empty neighboring sites that cannot be occupied by $i$, with the remaining neighbors being occupied. 

We define the {\it zero-switch blocked persistence} $\phi_b^{(0)}(t)$ as the fraction of particles that have been blocked up to time $t$ \emph{because} their blocking set has remained the same throughout the interval $(0,t)$. This zero-switch blocked persistence reads:
\begin{multline}
\label{eq:zeroSwitchKA}
\phi_b^{(0)}(t)=\rho \sum_{j=m+1}^z\, \binom{z}{j}\, Y^j(t)\, (1-Y(t))^{z-j} +\\ +\rho \sum_{j=0}^m \binom{z}{j}\, Y^j(t)\, B^{z-j}(t).
\end{multline}
For $z=4, m=2$ the possible contributions entering Eq.~\eqref{eq:zeroSwitchKA} can be represented graphically as:
\begin{equation}
\label{eq:diagPhi0}
\begin{split}
\phi_b^{(0)}(t)=&
\begin{gathered}
\scalebox{0.5}{
\begin{tikzpicture}
\node[label=below:{}, shape=circle, scale=1,fill=black!28,draw] (A) at (0,0) [right] {};
\draw[black,thick] (A) -- +(135:3\nodeDist);
\draw[black,thick] (A) -- +(45:3\nodeDist);
\draw[black,thick] (A) -- +(-45:3\nodeDist);
\draw[black,thick] (A) -- +(225:3\nodeDist);
\end{tikzpicture}
}
\end{gathered}
\quad+\quad
\begin{gathered}
\scalebox{0.5}{
\begin{tikzpicture}
\node[label=below:{}, shape=circle,semithick, scale=1,fill=black!28,draw] (A) at (0,0) [right] {};
\draw[black,thick] (A) -- +(135:3\nodeDist);
\draw[densely dashed,thick] (A) -- +(45:3\nodeDist);
\draw[black,thick] (A) -- +(-45:3\nodeDist);
\draw[black,thick] (A) -- +(225:3\nodeDist);
\end{tikzpicture}
}
\end{gathered}
+\\ & +\begin{gathered}
\scalebox{0.5}{
\begin{tikzpicture}
\node[label=below:{}, shape=circle,semithick, scale=1,fill=black!28,draw] (A) at (0,0) [right] {};
\draw[black,thick,double distance=2pt] (A) -- +(135:3\nodeDist);
\draw[black,thick,double distance=2pt] (A) -- +(45:3\nodeDist);
\draw[black,thick,double distance=2pt] (A) -- +(-45:3\nodeDist);
\draw[black,thick,double distance=2pt] (A) -- +(225:3\nodeDist);
\end{tikzpicture}
}
\end{gathered}
+\begin{gathered}
\scalebox{0.5}{
\begin{tikzpicture}
\node[label=below:{}, shape=circle,semithick, scale=1,fill=black!28,draw] (A) at (0,0) [right] {};
\draw[black,thick,double distance=2pt] (A) -- +(135:3\nodeDist);
\draw[black,thick,double distance=2pt] (A) -- +(45:3\nodeDist);
\draw[black,thick,double distance=2pt] (A) -- +(-45:3\nodeDist);
\draw[black,thick] (A) -- +(225:3\nodeDist);
\end{tikzpicture}
}
\end{gathered}
+\begin{gathered}
\scalebox{0.5}{
\begin{tikzpicture}
\node[label=below:{}, shape=circle,semithick, scale=1,fill=black!28,draw] (A) at (0,0) [right] {};
\draw[black,thick,double distance=2pt] (A) -- +(135:3\nodeDist);
\draw[black,thick,double distance=2pt] (A) -- +(45:3\nodeDist);
\draw[black,thick] (A) -- +(-45:3\nodeDist);
\draw[black,thick] (A) -- +(225:3\nodeDist);
\end{tikzpicture}
}
\end{gathered}
\end{split}
\end{equation}
where the full lines represent the neighbors of the blocked particle (circle) which have always remained occupied at all times less than $t$, the double lines represent empty neighbors that cannot be occupied, and the dashed line represents an unconditioned neighbor. Therefore, the first term in Eq.~\eqref{eq:zeroSwitchKA}, which takes into account the case in which there are at least $m+1$ occupied neighbors, is represented by the first two diagrams. The second in Eq.~\eqref{eq:zeroSwitchKA}, taking into account the case in which there are at least $z-m$ empty neighbors that cannot be occupied, and the remaining neighbors are occupied, is represented by the remaining three diagrams. Note that taking the infinite-time limit, Eq.~\eqref{eq:zeroSwitchKA} reduces to the equation \eqref{eq:defphiKA} for the plateau value of the persistence. We define analogously the zero-switch blocked cavity persistences $Y_b^{(0)}(t), B_b^{(0)}(t), F_b^{(0)}(t)$, and we use the notation:
\begin{equation}
\boldsymbol{\delta \hat{\phi}}_b^{(0)}(t)=\big(\delta Y_b^{(0)}(t),\delta B_b^{(0)}(t),\delta F_b^{(0)}(t)\big).
\end{equation}

We also define the {\it one-switch blocked persistence} $\phi_b^{(1)}(t)$ as the fraction of blocked particles not counted by $\phi_b^{(0)}(t)$ s.t.\ their blocking set is composed by $m$ neighbors that have always been occupied, and a so-called switching couple of neighbors. A switching couple is composed by a first neighboring site which has been occupied up to some time $t'$, and a second neighbor, that has been occupied in $(t'',t)$, where $0<t''<t'<t$. Note that $t''>0$, because the contribution $t''=0$ is already taken into account by $\phi_b^{(0)}(t)$ (for example, in the case $z=4,m=2$, the term with $t''=0$ is given by the second diagram in Eq.~\eqref{eq:diagPhi0}). Also $\phi_b^{(1)}(t)$ can be represented graphically. For $z=4,m=2$, we have:
\begin{equation}
\phi_b^{(1)}(t)=
\begin{gathered}
\scalebox{0.5}{
\begin{tikzpicture}
\node[label=below:{}, shape=circle,semithick, scale=1,fill=black!28,draw] (A) at (0,0) [right] {};
\draw[densely dashed,thick] (A) -- +(135:1.5\nodeDist);
\path (A) ++(135:1.5\nodeDist) coordinate (A2);
\draw[black,thick] (A2) -- +(135:1.5\nodeDist);
\draw[densely dashed,thick] (A) -- +(135:2\nodeDist);
\draw[densely dashed,thick] (A) -- +(45:3\nodeDist);
\draw[black,thick] (A) -- +(45:1.5\nodeDist);
\draw[black,thick] (A) -- +(-45:3\nodeDist);
\draw[black,thick] (A) -- +(225:3\nodeDist);
\end{tikzpicture}
}
\end{gathered}
.
\label{eq:phi1graph}
\end{equation}
The top lines in the diagram \eqref{eq:phi1graph} represent the \emph{switching couple} of neighbors: the top right line corresponds to the neighbor that is occupied up to time $t'$, and the top left line to the neighbor that is occupied between $t''$ and $t$. We define analogously the one-switch blocked cavity persistences $Y_b^{(1)}(t), B_b^{(1)}(t), F_b^{(1)}(t)$, that we denote by:
\begin{equation}
\boldsymbol{\hat{\phi}}_b^{(1)}(t)=\big(Y_b^{(1)}(t), B_b^{(1)}(t),F_b^{(1)}(t)\big).
\end{equation}

Following \cite{perrupato2025theory}, the one-switch blocked persistence $\phi_b^{(1)}(t)$ can be expressed in terms of a cavity persistence as follows:
\begin{multline}
\label{eq:phib1}
\phi_b^{(1)}(t)=-\rho\,\binom{z}{m}\,(z-m)\,(z-m-1)\,Y(t)^m\times\\\times  (1-Y(t))^{z-m-2}\, \int_0^t dt' \frac{dY(t')}{dt'}(Y(t-t')-Y(t)).
\end{multline}
In fact, let us call $i_1$ and $i_2$ the switching couple of neighbors of an occupied site $i_0$ counted by $\phi_b^{(1)}(t)$. Recalling the definition of $Y(t)$, it follows that the probability that $i_1$ was occupied up to $t'$, and that it is vacated between $t'$ and $t'+dt'$, is given by $-(d Y(t')/dt')  dt'$. The total probability that $i_2$ is occupied between time $t''$ and $t$ with $0<t''<t'$ can be computed invoking the reversibility of the dynamics: it is equal to the probability that, starting at equilibrium at time $t$, and moving backward in time, the site is occupied up to time $t-t'$ but not up to time $t$. This leads to the factor $Y(t-t')-Y(t)$ in \eqref{eq:phib1}. As discussed above, we have to subtract $Y(t)$ because the case $t'=0$ (and then $t''=0$) leads to a contribution which is already taken into account by the zero-switch persistence. Therefore, the probability that the neighbors $i_1$ and $i_2$ are a switching couple up to time $t$ is obtained by integrating $- (d Y(t')/dt')\,(Y(t-t')-Y(t))$ over $t'$. In order to obtain $\phi_b^{(1)}(t)$, we have to multiply this probability by: i) a combinatorial factor $\binom{z}{m}\,(z-m)\,(z-m-1)$ counting all possible ways of realizing the blocking set in the one-switch case; ii) the probability $\rho$ that $i_0$ is occupied in the initial condition; iii) the probability $Y^m(t)$ that $m$ of the neighbors of $i_0$ not belonging to the switching couple are occupied at all times less than $t$; iv) the probability $(1-Y(t))^{z-m-2}$ that the remaining neighbors are not always occupied in $(0,t)$. As discussed later, analogous expressions hold for the one-switch blocked cavity persistences (see App.~\ref{app:tdyneq}).

In the next section, we use the quantities defined here to write a closed equation for $\phi(t)$. 

\subsection{Critical hierarchy}
\label{ssec:critHier}

\begin{table}
\centering
\begin{tabular}{|c|c||c|c|c|c|c|}
\hline
$z$ &  $m$  & $\rho_c$ & $\phi_{plat}$  & $\lambda$ & $a$ & $b$\\ \hline\hline
4 & 2 & 0.888793 & 0.788605 & 0.673328 & 0.337761 & 0.685369 \\ \hline
5 & 2 & 0.724831 & 0.32489 & 0.716273 & 0.320017 & 0.613758 \\ \hline
5 & 3 & 0.948964 & 0.942315 & 0.649708 & 0.346793 & 0.725438 \\ \hline
6 & 2 & 0.602788 & 0.11982 & 0.734512 & 0.311883 & 0.583671 \\ \hline
6 & 3 & 0.834769 & 0.65877 & 0.701751 & 0.326226 & 0.637831 \\ \hline
6 & 4 & 0.970598 & 0.969832 & 0.643283 & 0.34917 & 0.736444 \\ \hline
7 & 2 & 0.513688 & 0.0477438 & 0.744701 & 0.307162 & 0.566909 \\ \hline
7 & 3 & 0.730949 & 0.523061 & 0.723766 & 0.316723 & 0.601382 \\ \hline
7 & 4 & 0.886856 & 0.777713 & 0.702238 & 0.326021 & 0.637021 \\ \hline
7 & 5 & 0.980847 & 0.954341 & 0.643002 & 0.349274 & 0.736926 \\ \hline
8 & 2 & 0.446765 & 0.182959 & 0.75112 & 0.304117 & 0.556359 \\ \hline
8 & 3 & 0.645919 & 0.430747 & 0.737636 & 0.31045 & 0.578529 \\ \hline
8 & 4 & 0.80069 & 0.655943 & 0.722238 & 0.3174 & 0.603903 \\ \hline
8 & 5 & 0.916342 & 0.843202 & 0.712058 & 0.321843 & 0.620734 \\ \hline
8 & 6 & 0.986518 & 0.969399 & 0.645149 & 0.348483 & 0.733243 \\ \hline
\end{tabular}
\caption{Dynamical parameters of the KA model on the Bethe lattice with connectivity $z$ and facilitation $m$.}
\label{tab:zflKA}
\end{table}

\begin{figure}
\centering
\includegraphics[width=\columnwidth]{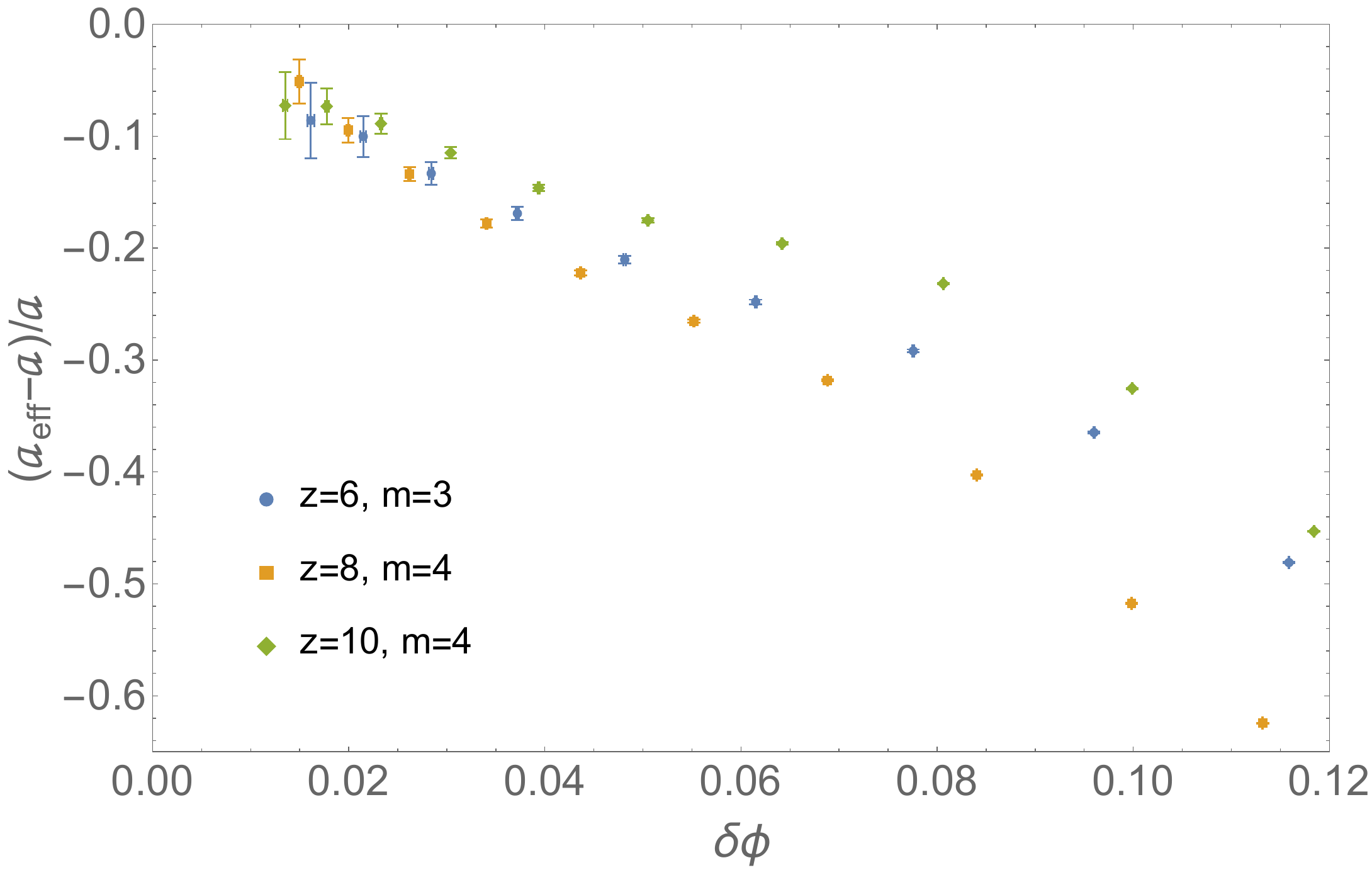}
\caption{Parametric plot of the relative shift of the effective
exponent $a_{eff}$ with respect to the analytical prediction $a$ vs.
the shift $\delta\phi$ from the plateau. From bottom right to top right: $z=8$ $m=4$, $z=6$ $m=3$ and $z=10$ $m=4$. Each point is obtained by performing numerical simulations on RRGs at different sizes ($10^5 <
N < 10^6$), and then extrapolating to infinite volume.\label{fig:logDerivKA}}
\end{figure}

\begin{figure}
\centering
\includegraphics[width=\columnwidth]{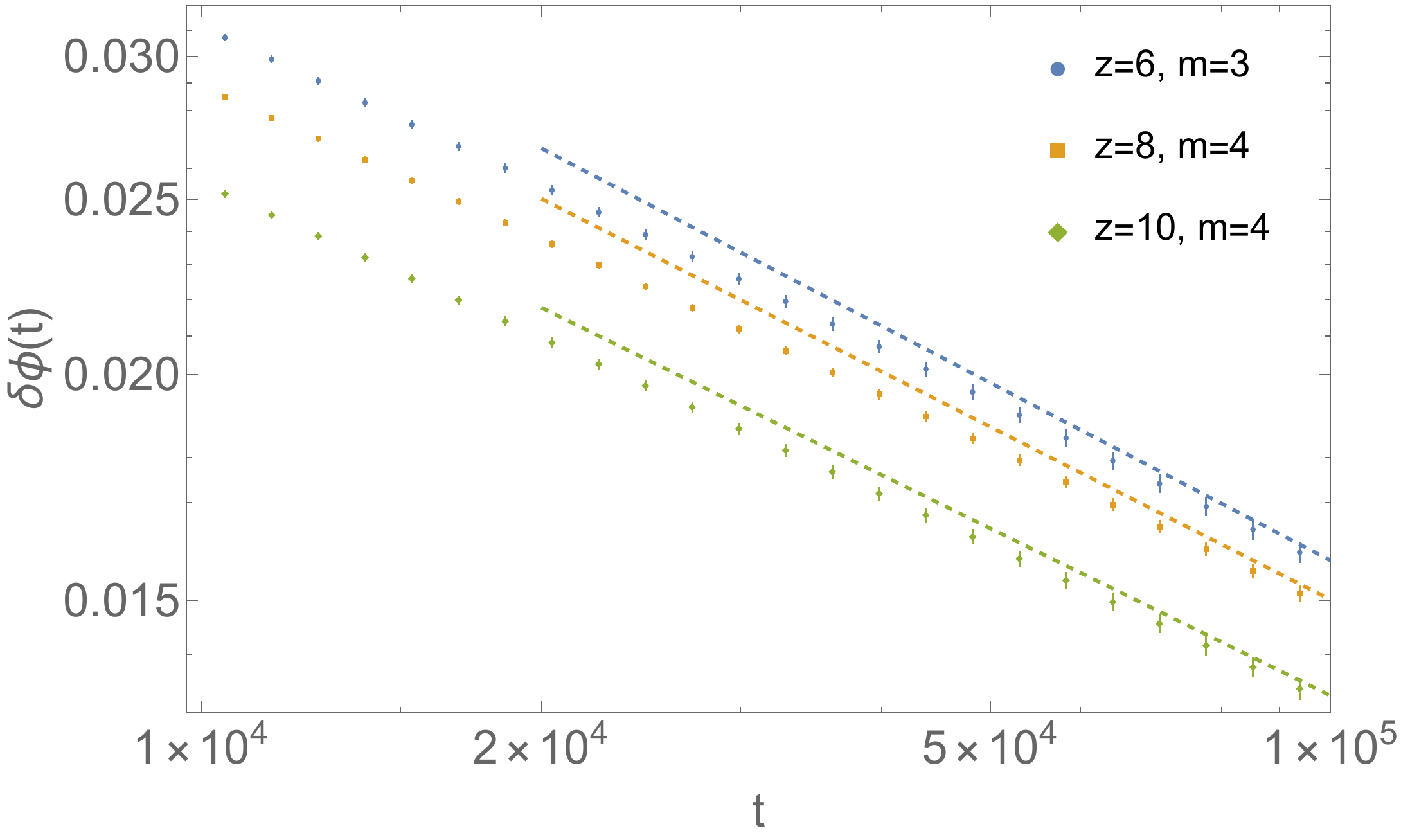}
\caption{Difference between the persistence and its plateau value vs $t$ at the critical density. From top right to bottom right: $z=8$ $m=4$, $z=6$ $m=3$ and $z=10$ $m=4$. Dashed lines correspond to curves proportional to $t^{-a_{z,m}}$, where $a_{z,m}$ are predicted analytically (see Table \ref{tab:zflKA}). The points represent numerical simulations on RRGs of size $N=4\times 10^5$.\label{fig:DeltaPhivsTDisc}}
\end{figure}

\begin{figure}
\centering
\includegraphics[width=0.95\columnwidth]{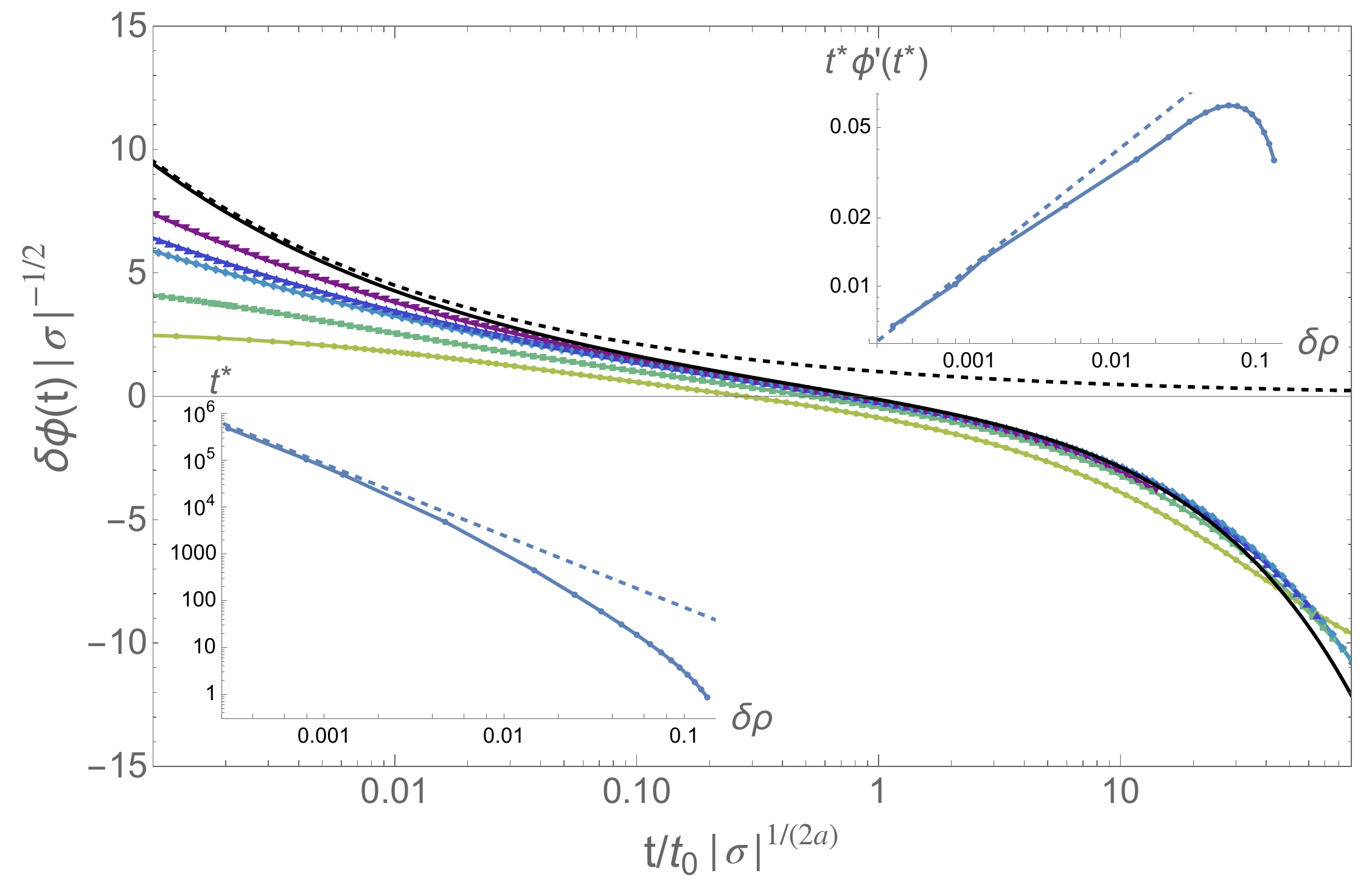}
\caption{Numerical check of the $\beta$-regime scaling law for $z=6$ and $m=3$ and $\rho\lesssim \rho_c$. The persistence in this regime is expected to be described by \eqref{eq:scalinglaw}, with $g(t)\equiv \delta\phi(t)=\phi(t)-\phi_{plat}$ and $\sigma=c_{z,m}\,(\rho-\rho_c)$, where the constant $c_{z,m}$ is predicted by our theory (see App.~\ref{app:tdyneq}). In particular, $c_{6,3}\approx 0.251$. The continuous black line represents the master function $f_{-}$ computed by solving numerically \eqref{MCTcrit} with $\sigma=-1$ and the value of $\lambda$ predicted in the case of $z=6$, $m=3$. The dashed line represents $t^{-a}$. The other curves correspond to numerical estimates of the persistence. From bottom left to top left $\rho=0.82, 0.83, 0.8335, 0.834, 0.8345$. The bottom left inset tests Eq.~\eqref{eq:scal1}. The dashed line represents $t_0\, \hat{t}\, \sigma^{-\frac{1}{2a}}$, with $t_0\approx 0.3$ fitted from the data at the critical point, and $a$ given by our analytical estimate. The top right inset tests Eq.~\eqref{eq:scal2}. The dashed line represents $\hat{t}\, f_{-}'(\hat{t})\, \sqrt{|\sigma|}$. Numerical simulations are obtained on RRGs with $N = 64 \times 10^5$ sites.\label{fig:scalMCT}}
\end{figure}

The blocked cavity persistence can be always written as the sum of zero-switch and one-switch persistences, plus an error $ \Delta \phi_b(t) >0 $ that counts all contributions to $\phi_b(t)$ other than $\phi_b^{(0)}(t)$ and $\phi_b^{(1)}(t)$:
\begin{equation}
\label{eq:phib12err}
\phi_b(t)=\phi_b^{(0)}(t)+\phi_b^{(1)}(t)+ \Delta \phi_b(t) \ .
\end{equation}
In FA \cite{perrupato2025theory} it is shown that in the $\beta$-regime a {\it hierarchy} between the different contributions emerges:
\beq
\label{eq:theHierarchy}
1 \gg \phi_b^{(0)}(t)-\phi_{plat} \gg \phi_b^{(1)}(t) \gg \Delta \phi_b(t) \ \ \ t \gg 1 \ .
\eeq
In particular, in the case of discontinuous transitions, it is found that $\phi_b^{(0)}(t)-\phi_{plat}\approx t^{-a}$, $\phi_b^{(1)}(t)\approx t^{-2a}$, $\Delta \phi_b(t)\approx t^{-3a}$ for some $a>0$ predicted by the theory. It is also shown that in order to compute the persistence $\phi(t)$ at the leading order in the $\beta$-regime, it is sufficient to take $\phi(t)\approx \phi_b(t)$ (see App.~\ref{app:EqPersBlockPers}), and to truncate \eqref{eq:phib12err} at the one-switch order:
\begin{equation}
\label{eq:trunchierarchy}
\phi(t)\approx \phi_b^{(0)}(t)+\phi_b^{(1)}(t).    
\end{equation}
Let us assume the validity of \eqref{eq:theHierarchy} and \eqref{eq:trunchierarchy}. We will later verify the consistency of this assumption by comparing the resulting predictions, e.g.\ the critical exponents and the scaling laws in the $\beta$-regime, with numerical simulations. 

Under \eqref{eq:theHierarchy}, the one-switch contribution $\phi_b^{(1)}(t)$ can be interpreted as a correction to the first term of \eqref{eq:zeroSwitchKA}, that corresponds to the case in which the occupied site is blocked because it has more than $m$ occupied neighbors. Note that we ignored one-switch corrections to the second term of \eqref{eq:zeroSwitchKA}, taking into account the case in which the occupied site is blocked because all its neighbors are blocked. This is the case because all neighbors must be blocked at all times, and the probability that a site switches from being blocked and occupied to being blocked and empty is negligible. 

From Eq.~\eqref{eq:trunchierarchy}, computing the critical behavior of $\phi(t)$ requires analyzing the cavity persistences, which determine both $\phi_b^{(0)}(t)$ and $\phi_b^{(1)}(t)$ through Eqs.~\eqref{eq:zeroSwitchKA} and \eqref{eq:phib1}, respectively. In order to do so, we proceed like for Eq.~\eqref{eq:trunchierarchy}, namely we express $Y(t), B(t), F(t)$ in terms of blocked persistences, and we truncate the latter at the one-switch term, which is expected to provide the leading behavior in the $\beta$-regime (see \cite{perrupato2025theory}). Therefore, we have:
\begin{equation}
\label{eq:01KAvec}
\boldsymbol{\delta \hat{\phi}}(t)\approx\boldsymbol{\delta\hat{\phi}}_b^{(0)}(t)+\boldsymbol{\hat{\phi}}_b^{(1)}(t).
\end{equation}
The one-switch terms $\boldsymbol{\hat{\phi}}_b^{(1)}(t)$ can be computed repeating the arguments leading to Eq.~\eqref{eq:phib1}. The complete expressions are written in App.~\ref{app:tdyneq}. The zero-switch blocked cavity persistences are exactly given by:
\begin{equation}
\label{eq:Yb0t}
Y_b^{(0)}(t) = \rho\, G^{(Y)}\big(Y(t),B(t)\big), 
\end{equation}
\begin{equation}
\label{eq:Bb0t}
B_b^{(0)}(t) = (1-\rho)\,G^{(B)}\big(F(t)\big),
\end{equation}
\begin{equation}
\label{eq:Fb0t}
F_b^{(0)}(t) = \rho\, G^{(F)}\big(Y(t),B(t)\big), 
\end{equation}
to be compared with Eqs.~\eqref{eq:KAcavY}, \eqref{eq:KAcavB} and \eqref{eq:KAcavF}. Close to the critical density $\rho_c$, i.e.\ for small $\delta \rho = \rho-\rho_c$, and on the timescale $\tau_{\beta}$ of the $\beta$-regime, we can expand Eqs.~\eqref{eq:Yb0t}, \eqref{eq:Bb0t} and \eqref{eq:Fb0t}, recalling that the cavity peristences are close to their plateau values. In this way, we obtain an expression of the form:
\begin{widetext}
\begin{equation}
\label{eq:dphi0KA}
\boldsymbol{\delta \hat{\phi}}_b^{(0)}(t)\approx \delta\rho\, \boldsymbol{\xi}+\boldsymbol{J}_{crit}\,\boldsymbol{\delta \hat{\phi}}(t)+\frac{1}{2}\left(\boldsymbol{\delta \hat{\phi}}^{\top}(t)\,\boldsymbol{H}^{(Y)}_{crit}\,\boldsymbol{\delta\hat{\phi}}(t),\boldsymbol{\delta \hat{\phi}}^{\top}(t)\,\boldsymbol{H}^{(B)}_{crit}\,\boldsymbol{\delta\hat{\phi}}(t),\boldsymbol{\delta \hat{\phi}}^{\top}(t)\,\boldsymbol{H}^{(F)}_{crit}\,\boldsymbol{\delta\hat{\phi}}(t)\right),
\end{equation}
\end{widetext}
where:  
\begin{enumerate}[label=\roman*)]
    \item $\boldsymbol{J}_{\text{crit}}$ is the (generally non-symmetric) Jacobian of the linearized system (see App.~\ref{app:cavityplat}), evaluated at $\rho=\rho_c$ and at the plateau values of the cavity persistences at the critical point;  
    \item $\boldsymbol{\xi}=\big(G^{(Y)}_{\text{crit}},-G^{(B)}_{\text{crit}},G^{(F)}_{\text{crit}}\big)$, with $G^{(Y)}_{\text{crit}}, G^{(B)}_{\text{crit}}, G^{(F)}_{\text{crit}}$ denoting the functions $G^{(Y)}, G^{(B)}, G^{(F)}$ evaluated at the same plateau values;  
    \item $\boldsymbol{H}_{\text{crit}}^{(Y)}, \boldsymbol{H}_{\text{crit}}^{(B)}, \boldsymbol{H}_{\text{crit}}^{(F)}$ are the Hessian matrices (second derivatives with respect to the cavity persistences) of $\rho_c G^{(Y)}$, $(1-\rho_c) G^{(B)}$, and $\rho_c G^{(F)}$, again taken at the plateau values at the critical point.
\end{enumerate}
The critical condition in Eq.~\eqref{eq:CritConddet} implies that the Jacobian $\boldsymbol{J}_{crit}$ admits a critical eigenvalue equal to one. We call, respectively, $\boldsymbol{v}^r$ and $\boldsymbol{v}^l$ the corresponding right and left critical eigenvectors. The dynamics in the $\beta$-regime is determined by the projection of the cavity persistences vector on the critical mode, since the other directions correspond to eigenvalues smaller than one, which imply an exponentially fast relaxation. For this reason, in the $\beta$-regime we write
\begin{equation}
\label{eq:shiftedCavpersRightEigKA}
\boldsymbol{\delta\hat{\phi}}(t)\approx g(t)\, \boldsymbol{v}^r,
\end{equation}
and we are interested in the projection of Eqs.~\eqref{eq:01KAvec} on $\boldsymbol{v}^l$, using the expansion \eqref{eq:dphi0KA} for $\boldsymbol{\delta \hat{\phi}}_b^{(0)}(t)$. By doing so, the linear terms in $g(t)$ cancel out, and we have to consider the $g(t)^2$ contributions that come both from the projection of \eqref{eq:dphi0KA} and from the one-switch term. This yields a quadratic equation for $g(t)$ in the form of \eqref{MCTcrit}, where $\sigma = c_{z,m}\, (\rho - \rho_c)$, and both $c_{z,m}$ and $\lambda$ are expressed as functions of $z$ and $m$, with explicit formulas provided in App.~\ref{app:tdyneq}. 

At this point, going back to the persistence $\phi(t)$, we have that in the $\beta$-regime, and for $1<m<z-1$ (discontinuous transitions) 
\begin{equation}
\label{eq:persistenceCavpersKA}
\delta\phi(t)\propto g(t),
\end{equation}
namely, the critical behavior of $\phi(t)$ is the same as that of the cavity persistences, like for FA in the case of discontinuous transitions. Eq.~\eqref{eq:persistenceCavpersKA} can be simply found substituting $Y+\delta Y(t)$ and $B+\delta B(t)$ into Eq.~\eqref{eq:zeroSwitchKA}, expanding at first order, and using Eq.~\eqref{eq:shiftedCavpersRightEigKA}. Therefore, also $\delta\phi(t)$ satisfies equation \eqref{MCTcrit} with the same value of $\lambda$. The analytical formula for $\lambda$ provided in App.~\ref{app:tdyneq} allows us to determine the critical exponents of the $\beta$-regime using Eq.~\eqref{lambdaMCT}. 

In table \ref{tab:zflKA} we display these critical exponents for all non-trivial values of $m$ up to $z=8$, excluded the cases with $m=1$, which correspond to continuous transitions (see Sec.~\ref{ssec:ContinuousModelsKA}). 

In Fig.~\ref{fig:logDerivKA} we study the effective exponent $a_{eff} \equiv  - d \ln \delta \phi / d \ln t$, which converges to the actual exponent $a$ at large times (small values of $\delta \phi$). 

In Fig.~\ref{fig:DeltaPhivsTDisc}, we compare the time-dependent difference between the persistence and its plateau value at the critical point, as obtained from numerical experiments, with our analytical predictions for the critical exponents for various values of $z$ and $m$. 

In Fig.~\ref{fig:scalMCT} we study the scaling laws of the $\beta$-regime, corresponding to timescales where $\phi(t)$ remains close to the plateau value. In order to do so, we consider for densities $\rho\lesssim \rho_c$ the time $t^*$ at which $\delta\phi(t^*)=0$, which can be measured in numerical experiments. Starting from Eq.~\eqref{eq:scalinglaw}, if $\hat{t}$ satisfies $f_{-}(\hat{t})=0$, then 
\begin{equation}
\label{eq:scal1}
t^*= \hat{t}\, \tau_{\beta}=\hat{t}\,t_0\, \sigma^{-\frac{1}{2a}},
\end{equation}
and
\begin{equation}
\label{eq:scal2}
t^*\,\delta\phi'(t^*)=\hat{t}\, f_{-}'(\hat{t})\, |\sigma|^{1/2},
\end{equation}
where $t_0$, the microscopic timescale that appears in the initial condition of the MCT equation \eqref{MCTcrit}, is not fixed by $\lambda$ and $\sigma$, and has to be fitted from the numerical data. In Fig.~\ref{fig:scalMCT} we test \eqref{eq:scal1}, and \eqref{eq:scal2}, and we study the critical collapse of $\delta\phi(\tau_{\beta}\,t)|\delta\rho|^{-1/2}$, obtained from numerical simulations, onto the master function $f_{-}(t)$, as predicted by Eq.~\eqref{eq:scalinglaw}. To compute the master function $f_{-}(t)$, we numerically solve Eq.~\eqref{MCTcrit} using the library provided in \cite{pihlajamaa2023modecouplingtheory}.

\subsection{Continuous models}
\label{ssec:ContinuousModelsKA}

\begin{figure}
\centering
\includegraphics[width=\columnwidth]{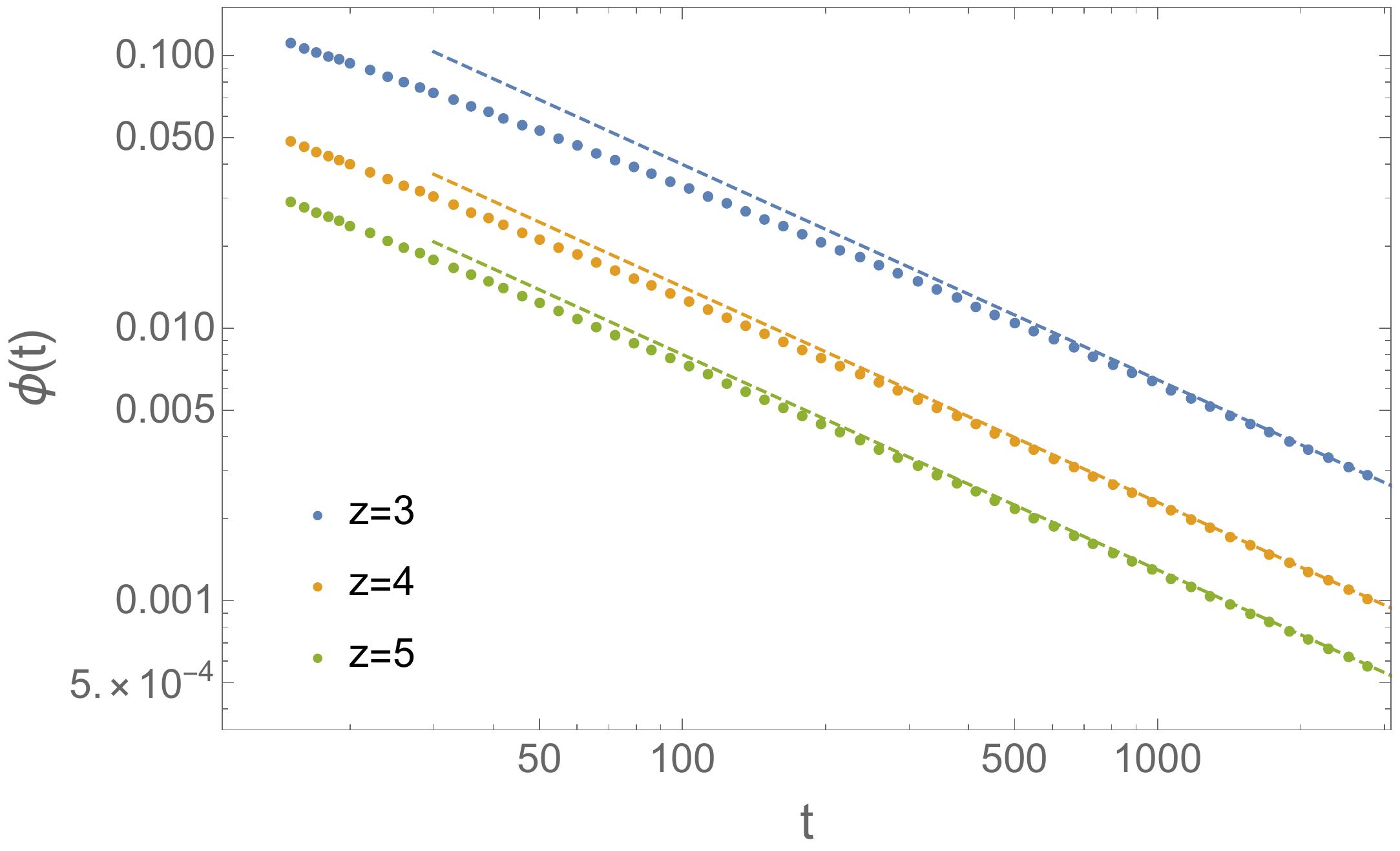}
\caption{Persistence in KA continuous models ($m=1$) on RRGs. From top to bottom: $z=3,4,5$.  The dashed lines correspond to $C_z t^{-2 a}$, where $a=0.395263$ is the analytical prediction, and $C_3\approx 1.52$, $C_4\approx 0.54$ and $C_5\approx 0.305$. Numerical data correspond to the average of $20$ samples of size $N=64\times 10^6$. \label{fig:continuiKA}}
\end{figure}

For $m=1$, the transition is continuous and occurs at $\rho_c=1/(z-1)$. In this case, the equations for $Y_b^{(0)}(t)$ and $Y_b^{(1)}(t)$ can be closed on $Y(t)$ at criticality. This fact simplifies the analysis of the previous section. In App.~\ref{app:Continuous} we show that for continuous transitions, independently of the value of the connectivity, $Y(t)$ decays as $t^{-a}$ with $\lambda = 1/2 \, \rightarrow  \  a=0.395263$, and $\phi(t)$ is quadratic in $Y(t)$, implying that {\it its dynamic exponent is doubled}: $\phi(t) \, \approx \, z\, Y^2(t)/2 \, \propto 1/t^{2 a}$. The same holds for continuous FA \cite{perrupato2025theory}. In Fig.~\ref{fig:continuiKA} we show the persistence for connectivity $z=3,4,5$, confirming the prediction that the exponent does not depend on $z$.

\subsection{$\alpha$ regime}
\label{sec:alphare}

\begin{figure}
\centering
\includegraphics[width=\columnwidth]{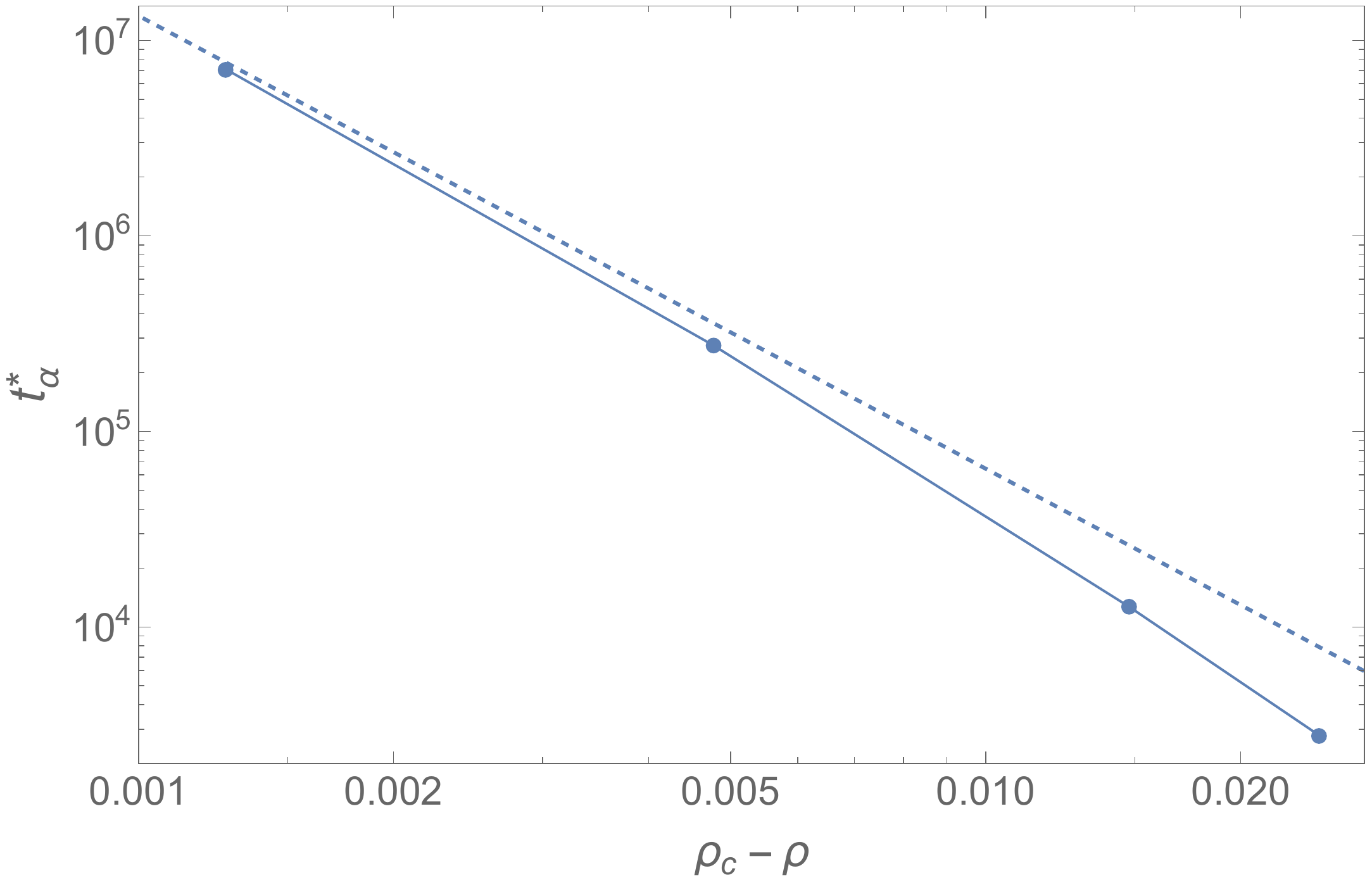}
\caption{Estimate of the $\alpha$ timescale $\tau_{\alpha}$, obtained computing the time $t^{*}_{\alpha}$ at which $\phi(t)=C$ as a function of $\rho_c-\rho$, where $C<\phi_{plat}$ is a constant. Here we choose $C=0.45$. If the scaling-law \eqref{eq:AlphaScalingLaw} is verified, then $t^{*}_{\alpha}\propto \tau_{\alpha}$. Data correspond to the case $z=6,m=3$ and are obtained on graphs with $N=64\times 10^5$ sites. The dashed line is proportional to $(\rho_c-\rho)^{-\gamma}$, where $\gamma= 2.31659\dots$ is the one predicted by our theory for $z=6,m=3$ (see Table \ref{tab:zflKA}).\label{fig:alphaRegimeScaling}}
\end{figure}

\begin{figure}
\centering
\includegraphics[width=\columnwidth]{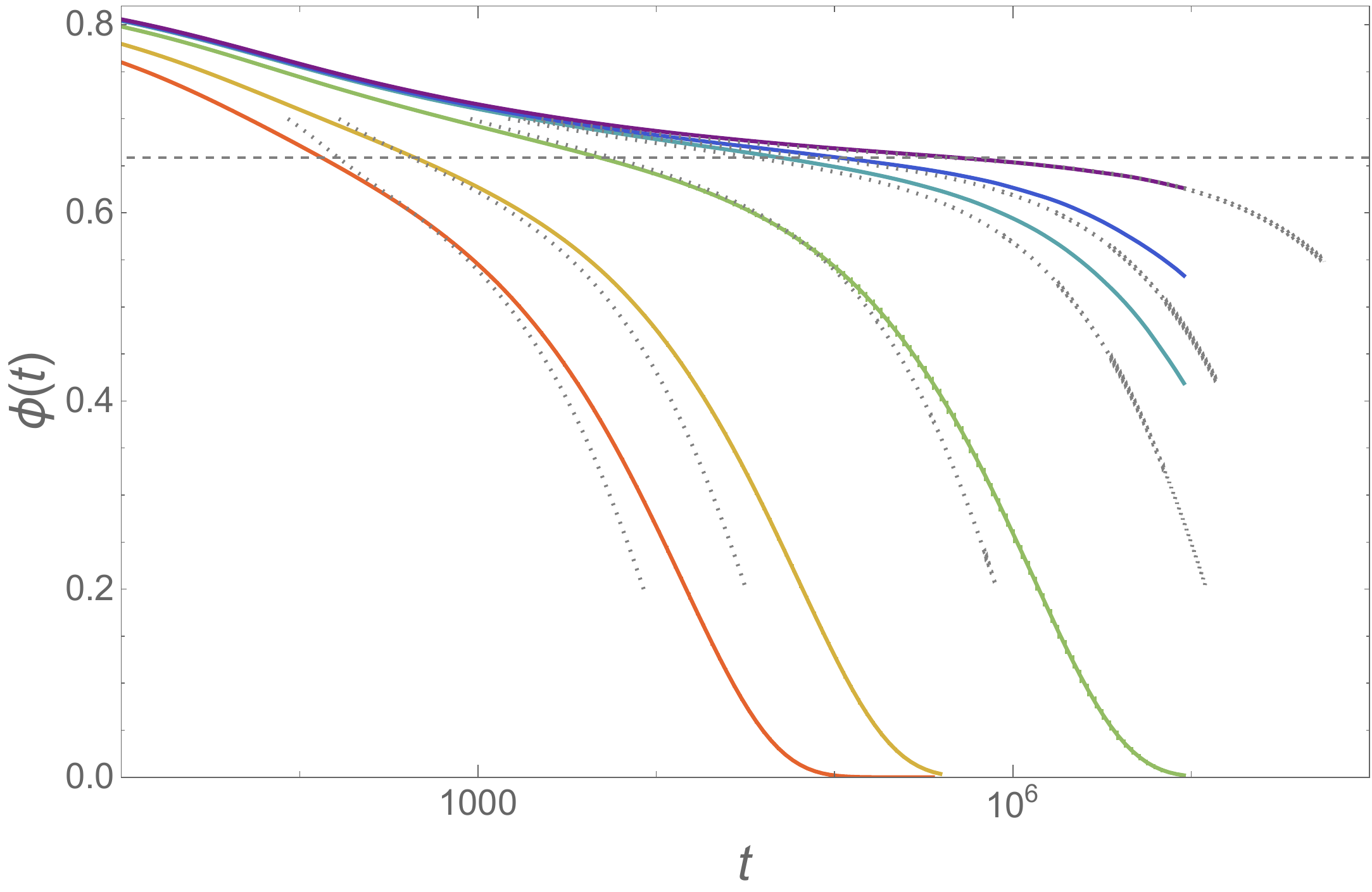}
\caption{Comparison between the persistence derived from the approximate scaling function of the $\alpha$-regime (dotted gray lines), and numerical data (continuous lines) for $\rho=0.81,0.82,0.83,0.8335,0.834,0.8345$. The scaling function, computed from the non-perturbative approximation obtained closing Eq.~\eqref{eq:01KAvec} with the complete expressions of $\boldsymbol{\hat{\phi}}_b^{(0)}(t)$ and $\boldsymbol{\hat{\phi}}_b^{(1)}(t)$ (see section \ref{sec:alphare}), reproduces qualitatively the behavior of the data. Numerical simulations are obtained on graphs with $N = 64 \times 10^5$ sites. \label{fig:alphaRegimeClosure}}
\end{figure}

The dynamical hierarchy leading to equation (\ref{MCTcrit}) is only valid in the $\beta$ regime, corresponding to timescales  where $\phi(t)$ remains close to the plateau value. 
Nonetheless, from the equation itself one can deduce, as we mentioned below Eq.~\eqref{MCTcrit}, not only the scaling law of the $\beta$ regime $\tau_\beta \propto |\sigma|^{-1/(2\,a)}$ but also the timescale of the $\alpha$ regime $\tau_\alpha$. The scaling function \eqref{eq:scalinglaw} implies that in the liquid phase $g(t)$ leaves the plateau with a power law behavior $-t^b$. The $\alpha$ regime can be identified as the time when $g(t)$ is no longer small and becomes $O(1)$.
This occurs when $(\tau_\alpha/\tau_{\beta})^b\, |\sigma|^{1/2}=O(1)$ leading to $\tau_\alpha \propto |\sigma|^{-\gamma}$ with $\gamma=1/(2 a)+1/(2 b)$. See Figure \ref{fig:alphaRegimeScaling} for a numerical check of the scaling $\tau_{\alpha}\propto |\sigma|^{-\gamma}$ in the case of KA, and \cite{sellitto2015} in the case of FA. 

In the $\alpha$ regime it can be argued that the following scaling law is obeyed:
\begin{equation}
\label{eq:AlphaScalingLaw}
 \phi(t) = C_{\alpha}(t/\tau_\alpha)   
\end{equation}
where the function $C_{\alpha}(x)$ is model-dependent at variance with the functions $f_{\pm}(x)$ that depend only on $\lambda$. 

The scaling function $C_{\alpha}(x)$ cannot be determined solely from Eq.~\eqref{MCTcrit}. This equation is quadratic in $g(t) = \delta \phi(t)$ and therefore is accurate as long as $\delta \phi$ is close to the plateau, i.e.\ on the time scale $\tau_\beta$. Instead in the $\alpha$ regime, by definition, $\delta \phi$ is $O(1)$ because the system leaves the plateau. As a consequence, terms of order $\delta \phi^3(t)$ and higher become important. These terms do not change the scaling laws of the critical exponents, i.e.\ $a$, $b$ and $\gamma$, but determine the model-dependent scaling function $C_{\alpha}(x)$. A non-perturbative approximation to the scaling function can be obtained closing Eq.~\eqref{eq:01KAvec} with the complete expressions of $\boldsymbol{\hat{\phi}}_b^{(0)}(t)$ (see Eqs.~\eqref{eq:Yb0t}, \eqref{eq:Bb0t}, \eqref{eq:Fb0t}) and $\hat{\phi}_b^{(1)}(t)$ (see Eqs.~\eqref{eq:eqDynKA1}, \eqref{eq:eqDynKA3} and \eqref{eq:eqDynKA2}). Note that this is different from what we were doing in the $\beta$ regime, where the expression for $\boldsymbol{\hat{\phi}}_b^{(0)}(t)$ is expanded close to its plateau value (see Eq.~\eqref{eq:dphi0KA}). This approximation yields a solution for $C_{\alpha}(t)$ that goes to zero at large times, because Eqs.~\eqref{eq:Yb0t}, \eqref{eq:Bb0t} and \eqref{eq:Fb0t} imply that $\hat{\phi}_b^{(0)}(t)=0$ if $\hat{\phi}_b(t)=0$, and $\hat{\phi}_b^{(1)}(t)$ tends to zero if $\hat{\phi}_b(t)$ tends to zero. We show this solution in Fig.~\ref{fig:alphaRegimeClosure}. Despite the fact that it satisfies $\hat{\phi}_b(\infty)=0$, it is not particularly accurate.

\section{Conclusions}  
\label{sec:concl}

In this work, we presented the analytical solution of the dynamics of the Kob-Andersen model on the Bethe lattice in the critical regime. Similarly to other models of supercooled liquids, the dynamics near the critical point exhibits a two-stage relaxation behavior, captured by the persistence function, which is an order parameter for the problem. Extending the combinatorial argument introduced in \cite{perrupato2025theory} for the Fredrickson-Andersen model (FA), we have shown that the persistence function obeys the very same critical equation of MCT. By means of numerical experiments, we have verified that the model displays indeed the whole MCT critical phenomenology, notably: i) the power-law divergences of the $\beta$ and $\alpha$ times scales as $\tau_\beta \propto |\delta \rho|^{-\frac{1}{2\,a}}$ and $\tau_\alpha \propto |\delta\rho|^{-\gamma}$, ii) $1/t^a$ approach to the plateau at the critical point and iii) the scaling law $ \phi(t)-\phi_{plat} \propto |\delta \rho|^{1/2}\, f_{\pm}(t/\tau_\beta)$ in the $\beta$ regime. Furthermore, the numerical data are in excellent quantitative agreement with our analytical predictions for the critical exponents. The theory has also been validated in the context of continuous transitions. 

We note that the critical mean-field dynamics of 1RSB spin glasses and of supercooled liquids in infinite dimensions is likewise governed by MCT equations of the same form as those derived for KA and FA. This demonstrates that a non-trivial free-energy landscape is not a necessary ingredient for the emergence of MCT-like critical phenomenology. Interestingly, although the transitions in these models are governed by the same equations, their physical properties differ. In spin glasses and in infinite-dimensional supercooled liquids, criticality is associated with the divergence of the static susceptibility inside the glassy states. By contrast, as shown in \cite{perrupato2024thermodynamics}, FA states are not critical, indicating that in general, criticality in the statics is not a necessary condition for criticality in dynamics.

 \begin{acknowledgments}
We acknowledge the financial support of the Simons Foundation (Grant No. 454949, Giorgio Parisi). 
\end{acknowledgments}

\appendix 

\section{Cavity equations for $Y, B$ and $F$}
\label{app:cavityplat}

Following \cite{toninelli2005cooperative}, we write the fixed-point equations for the plateau values of the cavity persistences:
\begin{widetext}
\begin{equation}
\label{eq:SEq1}
Y =  \rho\,\left(\sum_{j=m}^{z-1} \binom{z-1}{j} Y^j (1-Y)^{z-1-j} + \sum_{j=0}^{m-1} \binom{z-1}{j} Y^j B^{z-1-j}\right)\equiv \rho\, G^{(Y)}(Y,B), 
\end{equation}
\begin{equation}
\label{eq:SEq2}
B = (1-\rho) \sum_{j=m+1}^{z-1} \binom{z-1}{j} F^j (1-F)^{z-1-j}\equiv (1-\rho)\,G^{(B)}(F),
\end{equation}
\begin{equation}
\label{eq:SEq3}
F =  \rho\left(\sum_{j=m+1}^{z-1} \binom{z-1}{j} Y^j (1-Y)^{z-1-j} + \sum_{j=0}^{m} \binom{z-1}{j} Y^j B^{z-1-j}\right)\equiv \rho\, G^{(F)}(Y,B). 
\end{equation}
\end{widetext}
We now address the singularity of the Jacobian at the critical point (Eq.~\eqref{eq:CritConddet}). It is convenient to rewrite Eqs.~\eqref{eq:SEq1}, \eqref{eq:SEq2}, and \eqref{eq:SEq3} in the following form:
\begin{equation}
\boldsymbol{\mathcal{F}}(\boldsymbol{P},\rho) \equiv \boldsymbol{P} - \boldsymbol{\mathcal{G}}(\boldsymbol{P},\rho) = 0,
\end{equation}
where $\mathbf{P}=(Y,B,F)$ is the cavity persistence vector, and
\begin{equation}
\boldsymbol{\mathcal{G}}=\big(\rho\, G^{(Y)},\; (1-\rho)\, G^{(B)},\; \rho\, G^{(F)}\big).
\end{equation}
The map $\boldsymbol{\mathcal{F}}$ is continuous and differentiable. By the implicit function theorem, if for some $(\boldsymbol{P}^*,\rho^*)$ we have $\mathcal{F}(\boldsymbol{P}^*,\rho^*)=0$ and $\det\!\big(\mathds{1}-\boldsymbol{J}(\boldsymbol{P}^*,\rho^*)\big)\neq 0$, where $\boldsymbol{J}$ is the Jacobian matrix:
\begin{equation}
\big(\boldsymbol{J}(\boldsymbol{P},\rho)\big)_{ij}=\frac{\partial \boldsymbol{\mathcal{G}}_i(\boldsymbol{P},\rho)}{\partial \boldsymbol{P}_j},
\end{equation}
then there exists a locally unique smooth branch of solutions $\boldsymbol{P}(\rho)$ passing through $(\boldsymbol{P}^*,\rho^*)$. However, at the transition point, the glassy branch of solutions is created together with an unstable branch, and the two coincide exactly at $\rho=\rho_c$. This follows from the conservation of the Brouwer degree (see \cite{guckenheimer2013nonlinear}): for small $\rho$ there is a unique stable fixed point (index $+1$) \cite{toninelli2005cooperative}, and since the total degree is conserved, the emergence of an additional stable solution (index $+1$) necessarily entails the simultaneous appearance of an unstable one (index $-1$). The collision of these two branches at $\rho_c$ implies that local uniqueness is lost and, therefore, the Jacobian must become singular:
\begin{equation}
\det\!\big(\mathds{1}-\boldsymbol{J}_{\text{crit}}\big)=0,
\end{equation}
where $\boldsymbol{P}_{\text{crit}},\boldsymbol{J}_{\text{crit}}$ are, respectively, $\boldsymbol{P}$ and $\boldsymbol{J}$ computed at $\rho_c$.

\begin{figure}
\centering
\includegraphics[width=\columnwidth]{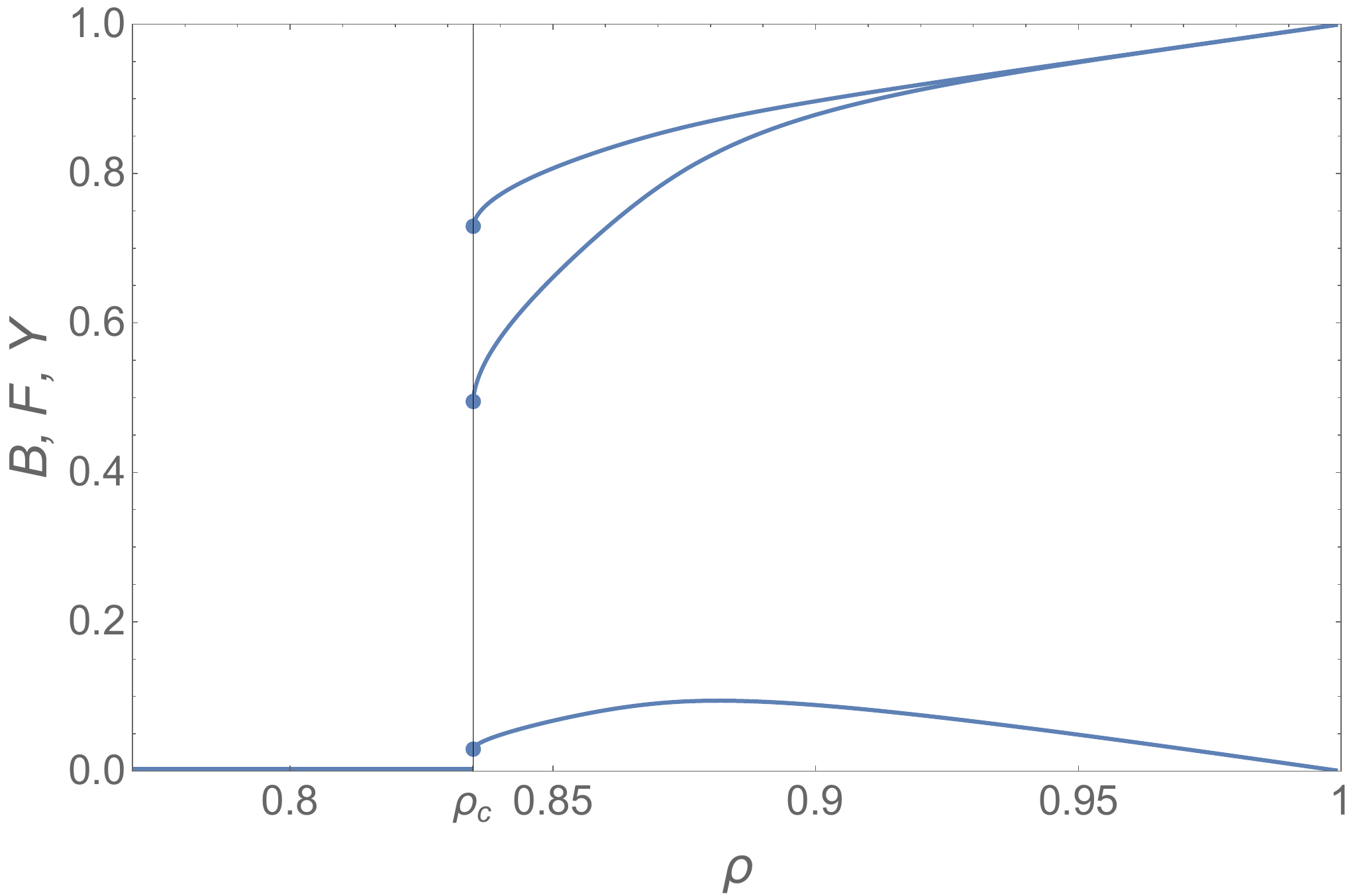}
\caption{Plateau values of the cavity persistences as a function of the density $\rho$, for $z=6$, $m=3$. From top to bottom $Y,F,B$. Upon increasing the density, at $\rho=\rho_c\approx 0.835$ the cavity persistences jump from a liquid solution $Y=B=F=0$, to a glassy solution associated with a non-zero value of the total persistence (see Fig.~\ref{fig:phiplatcav}).\label{fig:PlateauCavityPers}}
\end{figure}

\begin{figure}[b]
\centering
\includegraphics[width=\columnwidth]{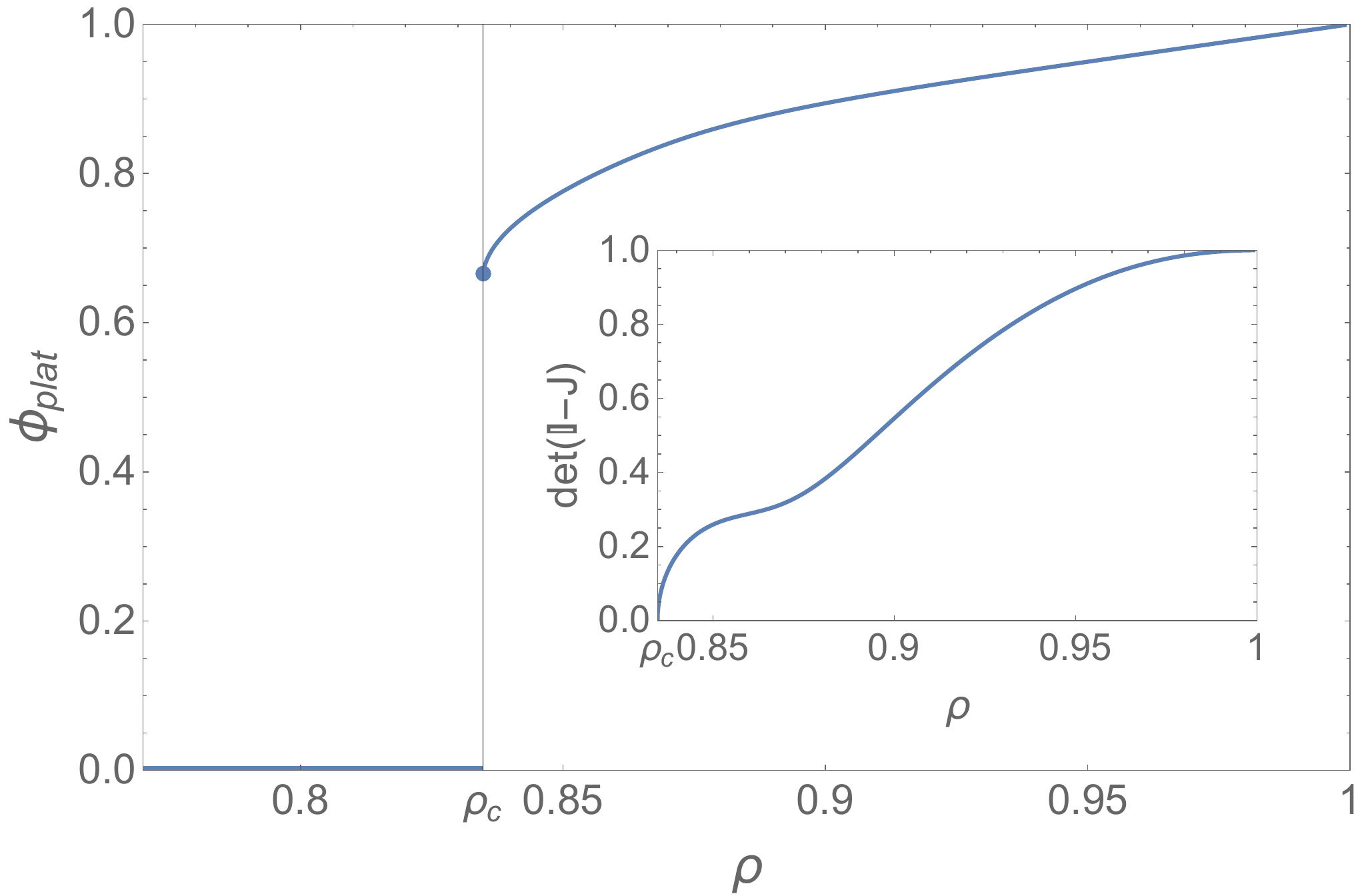}
\caption{Plateau value of the persistence as a function of the density $\rho$, for $z=6,\, m=3$. \emph{Inset}: value of $\det{\big(\mathds{1}-\boldsymbol{J}(\boldsymbol{P}(\rho),\rho)\big)}$ as a function of $\rho$, where $\boldsymbol{P}(\rho)=\left(Y(\rho),B(\rho),F(\rho)\right)$ is the vector of cavity persistences computed on the glassy fixed point at density $\rho$. Note that for $\rho\rightarrow \rho_c^+$ the determinant goes to zero.\label{fig:phiplatcav}}
\end{figure}

In Fig.~\ref{fig:PlateauCavityPers} we show the plateau value of $Y,B$ and $F$ as a function of the density $\rho$ for $z=6,m=3$, computed by solving the self-consistency Eqs.~\eqref{eq:SEq1}, \eqref{eq:SEq2} and \eqref{eq:SEq3}. In Fig.~\ref{fig:phiplatcav} we show for $z=6,\, m=3$ the plateau value of the persistence as a function of $\rho$, and $\det{\big(\mathds{1}-\boldsymbol{J}(\boldsymbol{P},\rho)\big)}$ computed on the glassy solution as a function of $\rho\in[\rho_c,1]$.

\section{Asymptotic equivalence between persistence and blocked persistence}
\label{app:EqPersBlockPers}
The argument is as follows. Let us start by noticing that the higher the number of times a particle is facilitated, the lower the probability that it does not move from its initial position. Now, because of the reversibility of the dynamics, if the particle was facilitated at some distant time in the past with probability one, it must have been facilitated many times at later times, leading to a vanishing probability that it did not move. In other words, we expect that once a particle becomes facilitated at time $t$, it will move with probability one after a finite time $t_{sw}$ that is short on the time scale of the critical dynamics. Therefore, the difference between the persistence $\phi(t)$ and the blocked persistence $\phi_b(t)$ is controlled by the fraction $\delta_{sw}(t)$ of the number of particles that become facilitated between times $t$ and $t+t_{sw}$, because these particle will typically start to move after $t_{sw}$. In formulas:
\begin{equation}
\label{eq:phibargum}
\phi(t)\approx\phi_b(t)+\delta_{sw}(t),
\end{equation}
where $\delta_{sw}$ is just given by: 
\beq
\delta_{sw}=\phi_b(t)-\phi_b(t+ t_{sw}) \approx - t_{sw}\frac{d \phi_b (t)}{dt}.
\eeq
For $m>1$ (discontinuous transitions), using the fact that in the $\beta$-regime $\phi(t)-\phi_{plat}\approx 1/t^{a}$, Eq.~\eqref{eq:phibargum} implies that for large times also $\phi_b(t)-\phi_{plat}\approx 1/t^{a}$, and in particular that their difference is $O(1/t^{a+1})$. As discussed in Sec.~\ref{ssec:critHier}, terms $O(1/t^{a+1})$ can be neglected for the characterization of the dynamics at the leading order. 

The same argument also applies to the continuous case ($m=1$), with the difference that $\phi(t)-\phi_{plat}\approx 1/t^{2a}$, which implies that the difference between persistence and blocked persistence is $O(1/t^{2a+1})$. 

\section{Dynamical equations}
\label{app:tdyneq}

As discussed in the main text, the critical dynamics is determined by the projection of the cavity persistences on the critical mode of the Jacobian $\boldsymbol{J}_{crit}$. For this reason, we write:
\begin{equation}
\label{eq:appParall}
\boldsymbol{\delta\hat{\phi}}(t)\approx g(t)\, \boldsymbol{v}^r,
\end{equation}
where $\boldsymbol{v}^r$ and $\boldsymbol{v}^l$ are, respectively, the right and the left critical eigenvectors. Let us denote the coordinates of $\boldsymbol{v}^r$ and $\boldsymbol{v}^l$, respectively, as $\boldsymbol{v}^r=(v^r_Y,v^r_B,v^r_F)$ and $\boldsymbol{v}^r=(v^l_Y,v^l_B,v^l_F)$. Projecting \eqref{eq:dphi0KA} on $\boldsymbol{v}^l$ we find 
\begin{equation}
\label{eq:projphi0}
\boldsymbol{v}^l\cdot\boldsymbol{\delta \hat{\phi}}_b^{(0)}(t)=g(t)+\delta\rho\,\boldsymbol{v}^l\cdot\boldsymbol{\xi}+C_{z,m}g(t)^2+\dots,
\end{equation}
where
\begin{multline}
C_{z,m}=\frac{v^l_Y}{2}(\boldsymbol{v}^{r})^T\boldsymbol{H}_{crit}^{(Y)}\boldsymbol{v}^{r}+\frac{v^l_B}{2}(\boldsymbol{v}^{r})^T\boldsymbol{H}_{crit}^{(B)}\boldsymbol{v}^{r}+\\+\frac{v^l_F}{2}(\boldsymbol{v}^{r})^T\boldsymbol{H}_{crit}^{(F)}\boldsymbol{v}^{r}.
\end{multline}
Following the same arguments of section \ref{sec:dynKAcav}, we can write the one-switch blocked cavity persistences in the following way:
\begin{widetext}
\begin{multline}
\label{eq:eqDynKA1}
Y_b^{(1)}(t) = - \rho\,\binom{z-1}{m-1} (z-m)(z-m-1) Y(t)^{m-1} (1-Y(t))^{z-m-2}  \int_0^t ds \frac{dY(s)}{ds}(Y(t-s)-Y(t)), 
\end{multline}
\begin{multline}
\label{eq:eqDynKA2}
B_b^{(1)}(t) = -(1-\rho)\,\binom{z-1}{m} (z-1-m)(z-2-m) F(t)^m (1-F(t))^{z-3-m} \int_0^t ds \frac{dF(s)}{ds}(F(t-s)-F(t)),
\end{multline}
\begin{multline}
\label{eq:eqDynKA3}
F_b^{(1)}(t) = -\rho\,\binom{z-1}{m} (z-1-m)(z-2-m) Y(t)^{m} (1-Y(t))^{z-m-3}  \int_0^t ds \frac{dY(s)}{ds}(Y(t-s)-Y(t)).
\end{multline}
\end{widetext}
Using Eq.~\eqref{eq:appParall}, and expanding at the leading order in $g(t)$, we get 
\begin{equation}
\label{eq:projphi1}
\boldsymbol{v}^{l}\cdot \boldsymbol{\hat{\phi}}^{(1)}_b=-D_{z,m}\,\int_0^t ds \frac{dg(s)}{ds}\big(g(t-s)-g(t)\big),
\end{equation}
where the coefficients $D_{z,m}$ are defined as follows:
\begin{widetext}
\begin{equation}
\begin{split}
D_{z,m}=&\rho_c\,\binom{z-1}{m-1}\, (z-m)(z-m-1)\, Y^{m-1}\, (1-Y)^{z-m-2}\,v_Y^l\,(v_Y^r)^2+\\
& +\rho_c\,\binom{z-1}{m}\, (z-1-m)(z-2-m)\, Y^{m}\, (1-Y)^{z-m-3}\,v_F^l\,(v_Y^r)^2+\\
&+(1-\rho_c)\,\binom{z-1}{m}\, (z-1-m)(z-2-m)\, F^m\, (1-F)^{z-3-m}\,v_B^l\,(v_F^r)^2.
\end{split}
\end{equation}
\end{widetext}
Therefore, substituting Eqs.~\eqref{eq:projphi1} and \eqref{eq:projphi0} into the projection of \eqref{eq:01KAvec} on $\boldsymbol{v}^l$, we get:
\begin{equation}
\label{eq:MCTform}
\frac{\boldsymbol{v}^l\cdot\xi}{D_{z,m}}\,\delta\rho=-\left(1+\frac{C_{z,m}}{D_{z,m}}\right)g(t)^2+\frac{d}{dt}\int_0^{t}ds\,g(s)\,g(t-s),
\end{equation}
where we used: 
\begin{multline}
\frac{d}{dt}\int_0^tds\,g(s)\,g(t-s)=\\=g^2(t)+\int_0^tds\frac{dg(s)}{ds}\big(g(t-s)-g(s)\big). 
\end{multline}
Note that Eq.~\eqref{eq:MCTform} has the same form of the MCT equation \eqref{MCTcrit}. The parameters $\lambda$ and $\sigma$ are given by: 
\begin{equation}
\lambda = 1+\frac{C_{z,m}}{D_{z,m}},\quad \sigma=\frac{\boldsymbol{v}^l\cdot\boldsymbol{\xi}}{D_{z,m}}\,\delta\rho.
\end{equation}

\section{Continuous transitions}
\label{app:Continuous}
For $m=1$ the transition is continuous, and the analysis of section \ref{sec:dynKAcav} simplifies, since to discuss the critical dynamics, the only relevant cavity persistence is $Y(t)$. Recall that in the continuous case $Y(t)$ tends to zero for large times. Expanding \eqref{eq:Yb0t} for small $Y(t)$ up to the second order at the critical point $\rho_c=1/(z-1)$, we find:
\begin{equation}
Y_b^{(0)}(t)=Y(t)-\frac{1}{2}(z - 2) Y(t)^2,
\end{equation}
and equation \eqref{eq:eqDynKA1} becomes
\begin{equation}
Y_b^{(1)}(t) = - (z-2)\int_0^t ds \frac{dY(s)}{ds}(Y(t-s)-Y(t)). 
\end{equation}
Therefore, using
\begin{equation}
Y(t)\approx Y_b^{(0)}(t)+Y_b^{(1)}(t),
\end{equation}
we find $\lambda = 1/2$, independently of $z$.
Now consider the persistence $\phi(t)$. The zero-switch blocked persistence $\phi_b^{(0)}(t)$ is given by: 
\begin{multline}
\label{eq:defphiKAtapp}
\phi_b^{(0)}(t) = \rho \sum_{j=m+1}^z \binom{z}{j} Y(t)^j (1-Y(t))^{z-j}+\\+\rho\sum_{j=0}^m \binom{z}{j} Y(t)^j B(t)^{z-j}.
\end{multline}
For $m=1$, at $\rho_c=1/(z-1)$, and close to the plateau, we have $\phi_b^{(0)}(t) \approx \frac{z}{2}\,Y(t)^2$, and the one-switch blocked persistence $\phi_b^{(1)}(t)$ (see the main text) is $O(Y(t)^3)$, therefore 
\begin{equation}
\phi(t) \approx \frac{z}{2}\,Y(t)^2.
\end{equation}

\newpage 
 
\bibliography{biblio}

\end{document}